\clearpage
\clearpage
\documentclass[12pt]{article}

\usepackage{times}
\usepackage{graphicx}

\newenvironment{sciabstract}{%
\pagenumbering{gobble}
\begin{quote} \bf}
{\end{quote}}

\usepackage{caption}
\newcounter{lastnote}

\title{Quantum criticality with two length scales} 

\author
{Hui Shao,$^{1,2,3}$ Wenan Guo,$^{1,4,\ast}$ Anders W. Sandvik$^{3,\ast}$\\
\\
\normalsize{$^{1}$Department of Physics, Beijing Normal University, Beijing 100875, China}\\
\normalsize{$^{2}$Beijing Computational Science Research Center, Beijing 100084, China} \\
\normalsize{$^{3}$Department of Physics, Boston University, Boston, Massachusetts 02215, USA}\\
\normalsize{$^{4}$State Key Laboratory of Theoretical Physics, Institute of Theoretical Physics,} \\
\normalsize{Chinese Academy of Sciences, Beijing 100190, China}
\\
\\
\normalsize{$^\ast$To whom correspondence should be addressed;}\\
\normalsize{E-mail: waguo@bnu.edu.cn, sandvik@bu.edu}
}

\date{}

\begin{document} 


\baselineskip24pt


\maketitle 


\begin{sciabstract}
The theory of “deconfined” quantum critical points describes phase transitions 
at temperature $T=0$ outside the standard paradigm, predicting continuous 
transformations between certain ordered states where conventional theory 
requires discontinuities. Numerous computer simulations have offered no proof 
of such transitions, however, instead finding deviations from expected scaling 
relations that were neither predicted by the DQC theory nor conform to standard scenarios. 
Here we show that this enigma can be resolved by introducing a critical 
scaling form with two divergent length scales. Simulations of a quantum magnet 
with antiferromagnetic and dimerized ground states confirm the form, proving 
a continuous transition with deconfined excitations and also explaining 
anomalous scaling at $T>0$. Our findings revise prevailing paradigms for 
quantum criticality, with potentially far-reaching implications for many 
strongly-correlated materials.
\end{sciabstract}

\clearpage


\paragraph*{Introduction}

\pagenumbering{arabic}

In analogy with classical phase transitions driven by thermal fluctuations, condensed matter systems can undergo drastic changes as parameters 
regulating quantum fluctuations are tuned at low temperatures. Some of these {\it quantum phase transitions} can be theoretically understood 
as rather straight-forward generalizations of thermal phase transitions \cite{fisher89,chubukov94}, where, in the conventional Landau-Ginzburg-Wilson 
(LGW) paradigm, states of matter are characterized by order parameters. Many strongly-correlated quantum materials seem to defy such a description, 
however, and call for new ideas.

A promising proposal is the theory of deconfined quantum critical (DQC) points in certain two-dimensional (2D) quantum 
magnets \cite{senthil04a,senthil04b}, where the order parameters of the antiferromagnetic (N\'eel) state and the competing dimerized state 
(the valence-bond-solid, VBS) are not fundamental variables but composites of fractional degrees of freedom carrying spin $S=1/2$. 
These {\it spinons} are condensed and confined, respectively, in the N\'eel and VBS state, and become deconfined at the DQC point
separating the two states. Establishing the applicability of the still controversial DQC scenario would be of great interest in condensed matter 
physics, where it may play an important role in strongly-correlated systems such as the cuprate superconductors \cite{kaul08}. There are also 
intriguing DQC analogues to quark confinement and other aspects of high-energy physics, e.g., an
emergent gauge field and the Higgs mechanism and boson \cite{huh13}. 

The DQC theory represents the culmination of a large body of field-theoretic works on VBS states and quantum phase transitions out of 
the N\'eel state \cite{read90,murthy90,read91,chubukov94,motrunich04}. The postulated SU($N$) symmetric non-compact (NC) CP$^{N-1}$ 
action can be solved when $N \to \infty$ \cite{metlitski08,kaul08,dyer15} but non-perturbative numerical simulations are required 
to study small $N$. The most natural physical realizations of the N\'eel--VBS transition for electronic SU(2) spins are frustrated quantum magnets 
\cite{read91}, which, however, are notoriously difficult to study numerically \cite{li12,gong14}. Other models were therefore pursued. 
In the $J$-$Q$ model \cite{sandvik07}, the Heisenberg exchange $J$ between $S=1/2$ spins is supplemented by a VBS-inducing
four-spin term $Q$ which is amenable to efficient quantum Monte Carlo (QMC) simulations
\cite{sandvik07,melko08,jiang08,sandvik10a,kaul11,harada13,chen13,block13,pujari15}. Although many results for the $J$-$Q$ model 
support the DQC scenario, it has not been possible to draw definite conclusions because of violations of expected scaling relations that 
affect many properties. Similar anomalies were later observed in three-dimensional loop \cite{nahum15} and dimer \cite{sreejith14} models,
which are also potential realizations of the DQC point. Simulations of the NCCP$^{1}$ action as well have been hard to reconcile with the theory
\cite{kuklov08,motrunich08,chen13}. 

One interpretation of the unusual scaling behaviors is that 
the transitions are first-order, as generally required within the LGW framework for order--order transitions where 
unrelated symmetries are broken. The DQC theory would then not apply to any of the systems studied so far, thus casting doubts on the entire 
concept \cite{jiang08,kuklov08,chen13}. In other interpretations the transition is continuous but unknown mechanisms cause strong corrections 
to scaling \cite{sandvik10a,motrunich08,bartosch13} or modify the scaling more fundamentally in some yet unexplained way 
\cite{kaul11,nahum15}. The enigmatic current state of affairs is well summed up in the recent Ref.~\cite{nahum15}.

Here we show that the DQC puzzle can be resolved based on a finite-size scaling Ansatz including the two divergent length scales of the 
theory---the standard correlation length $\xi$, which captures the growth of both order parameters ($\xi_{\rm N\acute{e}el} \propto \xi_{\rm VBS}$), 
and a faster-diverging length $\xi'$ associated with the thickness of VBS domain walls and spinon confinement (the size of a spinon bound state). 
We show that, contrary to past assumptions, $\xi'$ can govern the finite-size scaling even of magnetic properties that are sensitive only to 
$\xi$ in the thermodynamic limit. Our simulations of the $J$-$Q$ model at low temperatures and in the lowest $S=1$ (two-spinon) excited state 
demonstrate complete agreement with the two-length scaling hypothesis, with no other anomalous scaling corrections remaining.

\paragraph*{Finite-size scaling forms}

Consider first a system with a single divergent correlation length $\xi \propto |\delta|^{-\nu}$, where $\delta = g-g_c$ is the distance to a
phase transition driven by quantum fluctuations arising from non-commuting interactions controlled by $g$ at $T=0$ and $g_c$ is the critical
value of $g$. In finite-size scaling theory \cite{fsref}, for a system of linear size $L$ (volume $L^d$ in $d$ dimensions), close to 
$\delta=0$ a singular quantity $A$ takes the form
\begin{equation}
A(g,L)=L^{-\kappa/\nu}f(\delta L^{1/\nu},L^{-\omega}),
\label{aqlform1}
\end{equation}
where the exponents $\kappa,\nu,\omega$ are tied to the universality class, $\kappa$ also depends on $A$, and
$f$ approaches a constant when $\delta\to 0$ (up to scaling corrections $\propto L^{-\omega}$). We assume $T \propto L^{-z}$ (or, alternatively, 
$T=0$) so that scaling arguments depending  on $T$ have been eliminated. 

The form (\ref{aqlform1}) fails for some properties of the $J$-$Q$ model \cite{sandvik10a,kaul11,block13} and other DQC candidate systems 
\cite{kuklov08,nahum15,sreejith14}. A prominent example is the  spin stiffness, which for an infinite 2D system in the N\'eel phase should 
scale as  $\rho_s \propto \delta^{z\nu}$ with $z=1$ \cite{senthil04a,senthil04b,fisher89}. To eliminate the size dependence when $\delta\not=0$ and $L \to \infty$ 
in Eq.~(\ref{aqlform1}), we must have $\kappa=z\nu$ and $f(x,L^{-\omega}) \propto x^{z\nu}$ for large $x=\delta L^{1/\nu}$. 
Thus, $\rho_s(\delta=0,L) \propto L^{-z}$ and $L\rho_s$ should be constant when $L \to \infty$ if $z=1$. However, 
$L\rho_s(L)$ at criticality instead appears to diverge slowly \cite{jiang08,sandvik10a,chen13}. At first sight this might suggest 
$z < 1$, but other quantities, e.g., the magnetic susceptibility, instead behave as if $z > 1$ \cite{sandvik11}. 
Strong scaling corrections have been suggested as a way out of this paradox \cite{sandvik10a,kaul11,bartosch13}. 
Claims of a weak first-order transition have also persisted \cite{kuklov08,motrunich08,chen13}, though the continuous DQC scenario is supported
by the absence of any of the usual first-order signals, e.g., the Binder cumulant does not exhibit any negative peak \cite{sandvik10a,nahum15}. 

To explain the scaling anomalies phenomenologically, in the presence of a second length $\xi' \propto \delta^{-{\nu'}}$ in the VBS, we propose that 
Eq~(\ref{aqlform1}) should be replaced by the form
\begin{equation}
A(g,L)=L^{-\kappa/\tilde \nu}f(\delta L^{1/\nu},\delta L^{1/\nu'},L^{-\omega}),
\label{aqlform2}
\end{equation}
where, unlike what was assumed in the past, {\it $\tilde\nu$ is not necessarily the same as the exponent $\nu$ which governs the behavior of
most observables in the thermodynamic limit}. Instead, we show that the criticality in the $J$-$Q$ model generically demands $\tilde\nu=\nu'$. 

First assume $\tilde\nu=\nu$. The correct thermodynamic limit with $\kappa=z\nu$ for $\rho_s$ can then be obtained from Eq.~(\ref{aqlform2}) if
$f(x,y,L^{-\omega}) \propto x^{z\nu}$ for large $x=\delta L^{1/\nu}$, $y=\delta L^{1/\nu'}$ and, as before, $\rho_s(\delta=0,L) \propto L^{-z}$. 
This behavior can also be expressed using a scaling function where the second argument is the ratio of the two lengths; 
$\tilde f(\delta L^{1/\nu},L^{1/\nu'-1/\nu},L^{-\omega})$. If $\tilde f(\delta=0)$ is constant when 
$L \to \infty$, then $L^{1/\nu'-1/\nu}$ acts like just another irrelevant field, as in the standard scenario for dangerously irrelevant perturbations 
in classical clock models \cite{leonard15}. Our proposal is a different large-$L$ limit of Eq.~(\ref{aqlform2}) controlled by
$y=\delta L^{1/\nu'}$, which leads to concrete predictions of scaling anomalies.
In the case of the stiffness, the correct thermodynamic limit is obtained with $\tilde\nu=\nu'$ and $\kappa=z\nu$ if $f(x,y,L^{-\omega}) \propto y^{z\nu}$ 
for large $L$. Then $\rho_s(\delta=0) \propto L^{-z\nu/\nu'}$, which we can also obtain with $\tilde\nu=\nu$ and $\tilde f \propto L^{z(1-\nu/\nu')}$
for $\delta \to 0$. A function $\tilde f$ behaving as a power of $L$ for $\delta \to 0$ was implicitly suggested in Ref.~\cite{kaul11}, 
though with no specific form.

This alternative scaling behavior corresponds to $\xi \propto (\xi')^{\nu/\nu'}$ saturating at $\xi \propto L^{\nu/\nu'}$ when $\xi' \to L$ upon approaching 
the critical point, in contrast to the standard scenario where $\xi$ grows until it also reaches $L$ \cite{smnote}. The criticality at distances 
$r < L^{\nu/\nu'}$ is conventional, whereas $r > L^{\nu/\nu'}$ is governed by the unconventional power laws. Different behaviors for $r \ll L$ and 
$r \approx L$ were actually observed in the recent loop-model study \cite{nahum15} and a dangerously irrelevant field was proposed as a possible 
explanation, but with no quantitative predictions of the kind offered by our approach. The anomalous scaling law controlled by $\nu/\nu'$, which we 
confirm numerically below, is an unexpected feature of DQC physics and may also apply to other systems with two divergent lengths.

\paragraph*{Quantum Monte Carlo Results} 

The $J$-$Q$ model \cite{sandvik07} for $S=1/2$ spins is defined using singlet projectors $P_{ij} = 1/4 - {\bf S}_i \cdot {\bf S}_j$ as 
\begin{equation}
H = -J \sum_{\langle ij\rangle} P_{ij} -Q \sum_{\langle ijkl\rangle} P_{ij}P_{kl},
\label{jqham}
\end{equation}
where $\langle ij\rangle$ denotes nearest-neighbor sites on a periodic square lattice with $L^2$ sites and the pairs $ij$ and $kl$ in $\langle ijkl\rangle$ 
form horizontal and vertical edges of $2\times 2$ plaquettes. This Hamiltonian has all symmetries of the square lattice and the VBS ground state existing for 
$g=J/Q < g_c$ (with $g_c \approx 0.045$) is columnar, breaking the translational and $90^\circ$ rotational symmetries spontaneously. The N\'eel state for $g > g_c$ 
breaks the spin-rotation symmetry. 

We will study several quantities in the neighborhood of $g_c$. Although we have argued that the asymptotic $L\to \infty$ behavior when $\delta \not=0$ 
in Eq.~(\ref{aqlform2}) is controlled by the second argument of $f$, the critical finite-size scaling close to $\delta=0$ (when $\delta L^{1/\nu}$ is
of order $1$) can still be governed 
by the first argument \cite{smnote}. We will demonstrate that, depending on the quantity, either $\delta L^{1/\nu}$ or $\delta L^{1/\nu'}$ is the 
relevant argument, and, therefore, $\nu$ and $\nu'$ can be extracted using finite-size scaling with effectively single-parameter forms. 
We will do this for manifestly dimensionless quantities, $\kappa=0$ in Eq.~(\ref{aqlform2}), before testing the anomalous powers of $L$ 
in other quantities.

If the effective one-parameter scaling holds close to $g_c$, then Eq.~(\ref{aqlform2}) implies that $A(g,L_1)=A(g,L_2)$ at some $g=g^*(L_1,L_2)$ and 
a crossing-point analysis (Fisher's phenomenological renormalization) can be performed \cite{fsref}. For a $\kappa=0$ quantity, if $L_1=L$ and $L_2=rL$ 
with $r>1$ constant, a Taylor expansion of $f$ shows that the crossing points $g^*(L)$ approach $g_c$ as $g^*(L)-g_c \propto L^{-(1/\nu+\omega)}$ if 
$\nu$ is the relevant exponent (which we assume here for definiteness). $A^*=A(g^*)$ approaches its limit $A_c$ as 
$A^*(L)-A_c \propto L^{-\omega}$, and one can also show that
\begin{equation}
\frac{1}{\nu^*(L)} = \frac{1}{\ln(r)}\ln{ \left ( \frac{{\rm d}A(g,rL)/{\rm d}g}{{\rm d}A(g,L)/{\rm d}g}\right )_{g=g^*}}
\label{nustar1}
\end{equation}
converges to $1/\nu$ at the rate $L^{-\omega}$. In practice, simulation data can be generated on a grid of points close to the 
crossing values, with polynomials used for interpolation and derivatives. We present details and tests of such a scheme for the Ising model
in Supplementary Material.

\begin{figure}
\center{\includegraphics[width=10cm, clip]{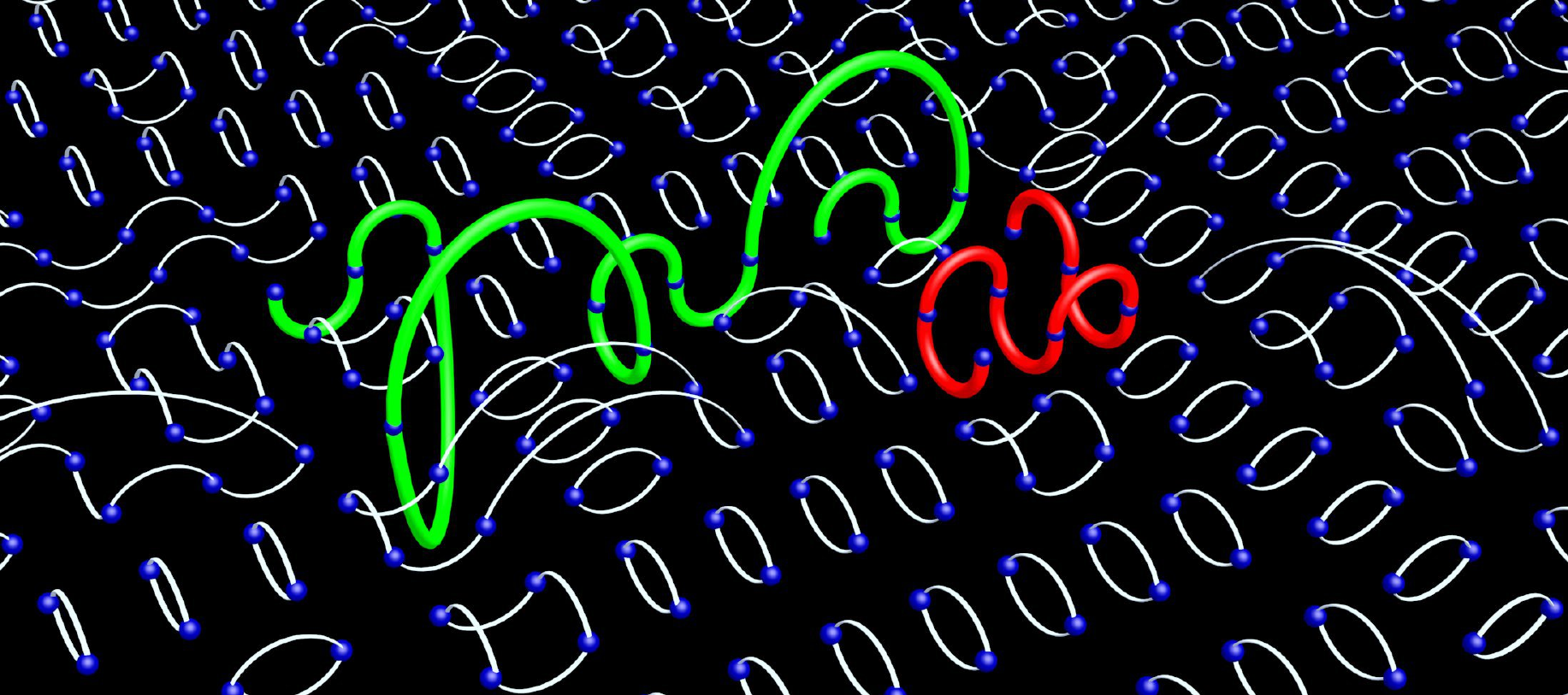}}
\vskip-1mm
\caption{A QMC transition graph representing \cite{tang11,banerjee10} a sampled overlap $\langle \psi_L|\psi_R\rangle$ of $S=1$ states, with 
two strings (spinons) in a background of loops formed by valence bonds. Arches above and below the plane represent $|\psi_R\rangle$ and
$\langle \psi_L|$, respectively.}
\label{fig0}
\vskip-3mm
\end{figure}
 
In the $S=1$ sector, spinon physics can be studied with projector QMC simulations in a basis of valence bonds (singlet pairs) 
and two unpaired spins \cite{tang11,banerjee10}. Previously the size of the spinon bound state in the $J$-$Q$ model was extrapolated to the 
thermodynamic limit \cite{tang13}, but the results were inconclusive as to the rate of divergence upon 
approaching the critical point. Here we study the critical finite-size behavior. We define the size $\Lambda$ of the spinon pair using the strings 
connecting the unpaired spins in valence-bond QMC simulations \cite{tang11,banerjee10}, as illustrated in Fig.~\ref{fig0} and further discussed
in Supplementary Material.

\begin{figure}[t]
\center{\includegraphics[width=12cm, clip]{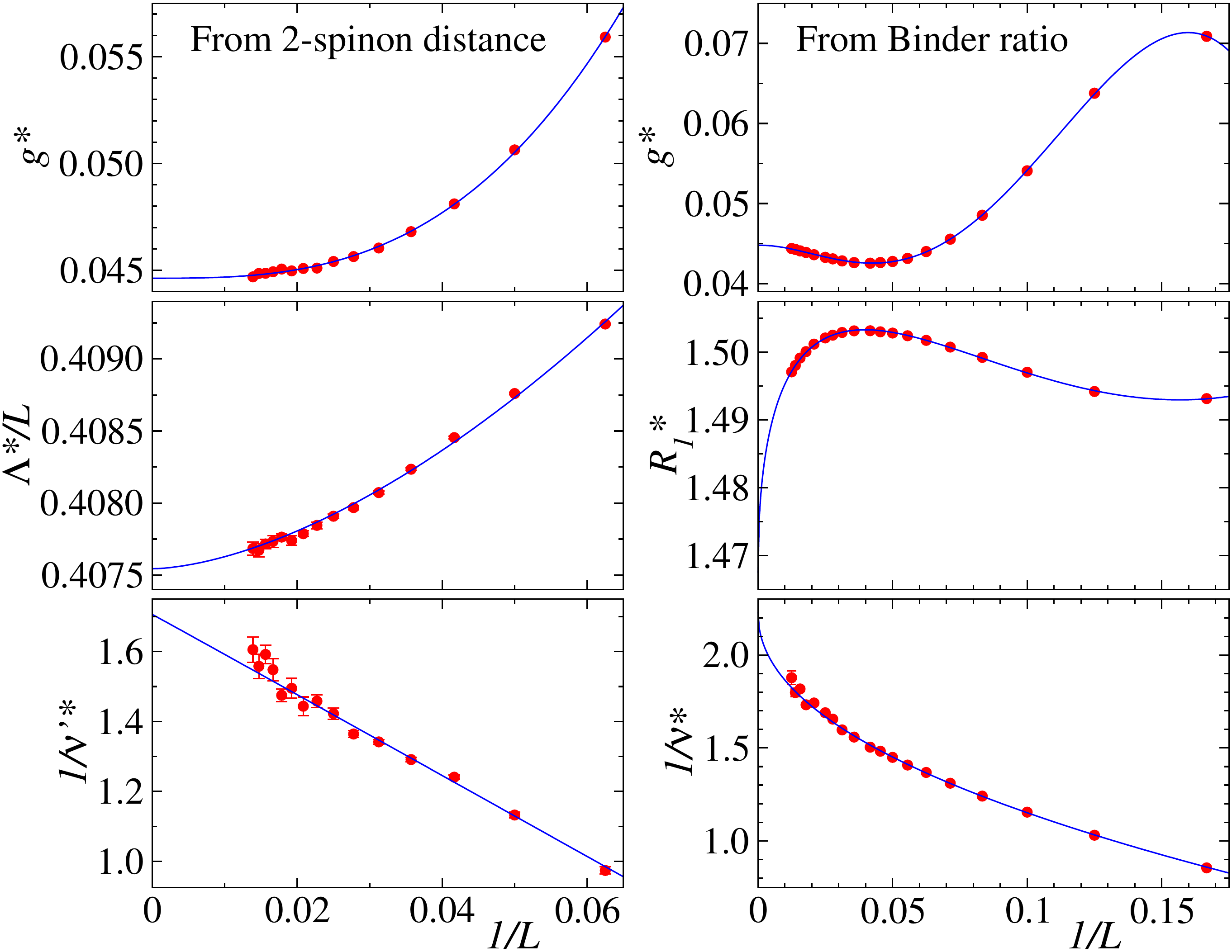}}
\vskip-1mm
\caption{Crossing-point analysis of $(L,2L)$ pairs for the size of the spinon bound state (left) and the Binder ratio (right). The
monotonic quantities are fitted with simple power-law corrections, while two subleading corrections were included in the fits of
the non-monotonic quantities.}
\label{fig1}
\vskip-3mm
\end{figure}

If $\Lambda(g) \propto \xi'(g)$ when $L \to \infty$, then $\Lambda(g_c) \propto L$ follows from our proposed limit of
Eq.~(\ref{aqlform2}). If $\Lambda$ manifestly probes only the longer length scale also in a finite system, which we will test, then
$\nu'$ is the exponent controlling the crossing points of $\Lambda/L$. Data and fits are presented in Fig.~\ref{fig1} (left). 
Unlike other quantities used previously to extract the critical point \cite{sandvik10a}, the drift of $g^*$ with $L$ is monotonic in this case and the 
convergence is rapid. All $L \ge 16$ points are consistent with the expected power-law correction, with $1/\nu'+\omega \approx 3.0$ 
and $g_c =0.04468(4)$, where the number in parenthesis indicates the statistical uncertainty (one standard deviation) in the preceding 
digit. The critical point agrees well with earlier estimates \cite{sandvik10a}. The crossing value $\Lambda^*/L$ also clearly converges
and a slope analysis according to Eq.~(\ref{nustar1}) gives $\nu' = 0.585(18)$.

Next, in Fig.~\ref{fig1} (right) we analyze a Binder ratio, 
defined with the $z$-component of the sublattice magnetization $m_{sz}$ as 
$R_1 = \langle m_{sz}^2\rangle/\langle |m_{sz}|\rangle^2$ and computed at $T = 1/L$ as in Ref.~\cite{sandvik10a}. Here the 
non-monotonic behavior of the crossing points necessitates several scaling corrections, unless only the largest sizes are used. In either case, 
the $L \to \infty$ behavior of $g^*$ is fully consistent with $g_c$ obtained from $\Lambda/L$. $R_{1}(g_c)$ has an uncertainty of over $1\%$ because 
of the small value of the correction exponent; $\omega\approx 0.4 \sim 0.5$. The slope estimator (\ref{nustar1}) of the exponent $1/\nu$ is 
monotonic and requires only a single $L^{-\omega}$ correction, also with a small exponent $\omega \approx 0.45$. The extrapolated exponent $\nu=0.446(8)$ is close 
to the value obtained recently for the loop model \cite{nahum15}. 

\begin{figure}
\center{\includegraphics[width=15cm, clip]{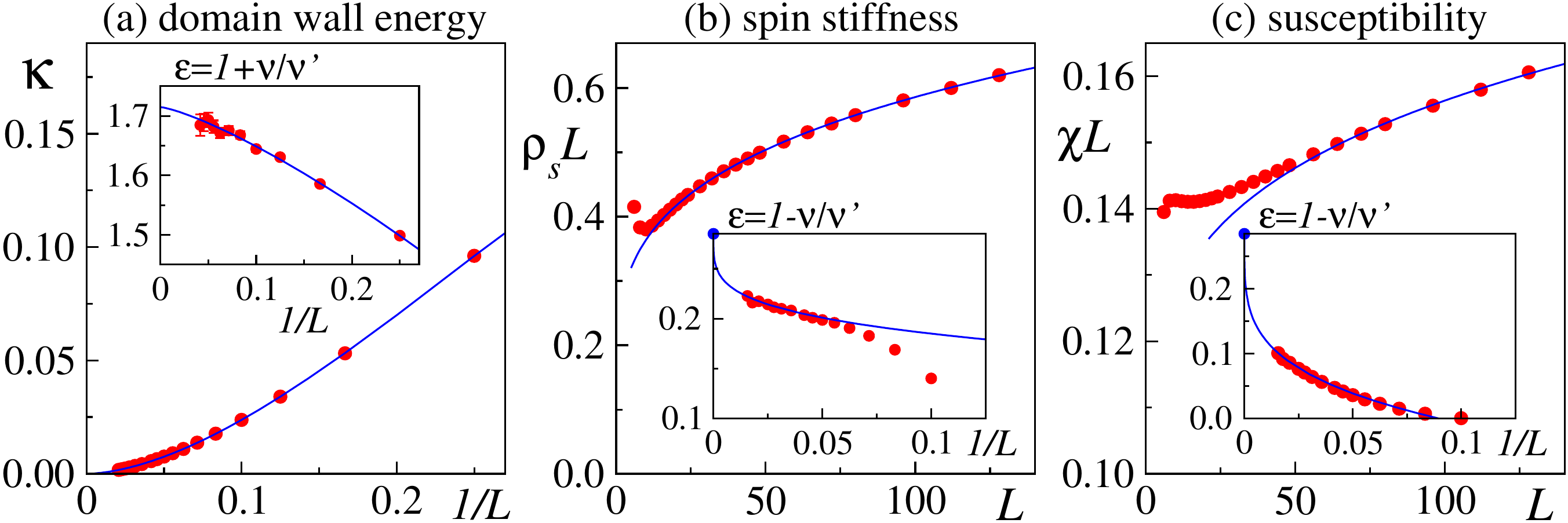}}
\vskip-1mm
\caption{Consistent anomalous critical scaling of different quantities $y$ at $J/Q=0.0447$. The insets show running exponents 
$\epsilon(L)=|\ln({y_{L}/y_{2L}})|/\ln(2)$ based on $(L,2L)$ data. In (a), a fit of $\epsilon(L)$ gave $1+\nu/\nu'=1.715$ for $L \to \infty$ 
and a correction $\propto L^{-1.2}$. In (b) and (c), $1-\nu/\nu'$ was fixed at the corresponding value $0.285$ and corrections 
$\propto L^{-\omega}$ with $\omega \approx 0.3$ were fitted to $\epsilon(L)$ for large systems. The same values of $\nu/\nu'$ and $\omega$
were used in fitted curves of the form $L^{1-\nu/\nu'}(a+bL^{-\omega})$ in the main (b) and (c) graphs.}
\label{fig3}
\vskip-3mm
\end{figure}

The above results support a non-trivial deconfinement process where the size of the bound state diverges faster than the conventional correlation length.
However, in the DQC theory the fundamental longer length scale $\xi'$ is the thickness of a VBS domain wall. It can be extracted from the domain wall energy
per unit length, $\kappa$, which in the thermodynamic limit should scale as $\kappa \propto (\xi\xi')^{-1}$ \cite{senthil04b}. In Supplementary Material 
we re-derive this form using a two-length scaling Ansatz and discuss simulations of domain walls in a 3D clock model 
and the $J$-$Q$ model. At criticality, in the conventional scenario (exemplified by the clock model) both $\xi$ and $\xi'$ saturate at $L$ and 
$\kappa \propto L^{-2}$. For the $J$-$Q$ model we instead find $\kappa \propto L^{-a}$ with $a = 1.715(15)$ for large $L$, as illustrated in 
Fig.~\ref{fig3}(a). Our interpretation of this unconventional scaling is that, when $\xi'$ saturates at $L$, $\xi$ also stops growing 
and remains at $\xi \propto L^{\nu/\nu'}$. Thus $\kappa \propto L^{-(1+\nu/\nu')}$ with $\nu/\nu'=a-1 = 0.715(15)$, which agrees reasonably well with 
$\nu/\nu'=0.76(3)$ obtained from the quantities in Fig.~\ref{fig1}. The large error bar on the latter ratio leaves open the possibility that the spinon 
confinement exponent is between $\nu$ and the domain-wall exponent $\nu'$ \cite{senthil04b}. 

We also calculated
the critical spin stiffness $\rho_s$ and susceptibility $\chi(k=2\pi/L)$ for the smallest wave-number $k=2\pi/L$, using $T=1/L$. Conventional quantum-critical 
scaling dictates that these quantities should decay as $1/L$ when $z=1$ \cite{chubukov94}. 
Instead, Figs.~\ref{fig3}(b) and (c) demonstrate clearly slower decays, 
$L\rho_s$ and $L\chi$ being slowly divergent, as had been found in earlier works as well\cite{jiang08,sandvik10a,sandvik11,kaul11,chen13}. The unconventional 
limit of the scaling function (\ref{aqlform2}) requires $L\rho_s$ and  $L\chi$ to diverge as $L^{1-\nu/\nu'}$. The behaviors are indeed consistent with this
form and $\nu/\nu' \approx 0.715$ extracted from $\kappa$, with a correction $\propto L^{-\omega}$ with a small $\omega$ (close to the correction for $R_1$ in 
Fig.~\ref{fig1}). The mutually consistent scaling behavior of the three different quantities lends strong support to the new type of criticality 
where the magnetic properties are not decoupled from the longer VBS length scale $\xi'$ for finite $L$. The results are incompatible with a 
first-order transition, where, $\kappa \to {\rm constant}$, $L\rho_s \to L$, $L\chi \to L$.

\paragraph*{Discussion}

We have shown that the effects of the larger divergent length scale $\xi'$ at the N\'eel--VBS transition are more dramatic than those caused by 
standard \cite{leonard15} dangerously-irrelevant perturbations, and we therefore propose the term {\it super dangerous} for this case. The universality
class, in the sense of the normal critical exponents in the thermodynamic limit at $T=0$, are not affected by such perturbations, but anomalous power 
laws of the system size appear generically in finite-size scaling. We have determined the value $\nu/\nu' \approx 0.72$ for the exponent ratio governing 
the anomalous scaling in the $J$-$Q$ spin model.

Loop and dimer models exhibit similar scaling anomalies \cite{nahum15,sreejith14} and it would be interesting to test the consistency between different 
quantities as we have done here. In simulations of the NCCP$^{1}$ action \cite{chen13,kuklov08,motrunich08} one would at first sight not expect 
any effects related to the longer DQC length scale, because the monopoles responsible for the VBS condensation are not present in the continuum 
theory \cite{senthil04a}. There could still be some other super dangerous operator present [see also Ref.~\cite{nahum15}], 
perhaps related to lattice regularization.

The consequences of our findings extend also to $T>0$ quantum criticality in the thermodynamic limit, because $1/T$ is the thickness of an equivalent 
system in the path integral formulation \cite{fisher89,chubukov94}. Anomalous finite-$T$ behaviors of the $J$-$Q$ model have already been observed 
\cite{sandvik10a,sandvik11}. For instance, the spin correlation length at $T>0$, which should be affected by deconfined spinons, grows faster than the normally 
expected form $\propto T^{-1}$ and the susceptibility vanishes slower than $T$. Remarkably, the asymptotic forms $\propto T^{-\nu'/\nu}$ and $T^{\nu/\nu'}$ 
can account for the respective behaviors (Fig.~S10 \cite{smnote}). Thus, we find a strong rationale to revise the experimentally most important tenet of
quantum criticality---the way scaling at the $T=0$ critical point is related to power laws in $T$ at $T>0$.

We conclude that quantum criticality with two divergent length scales is much richer than previously anticipated. Our findings may apply 
to a wide range of strongly-correlated quantum systems with more than one length scale and may help to resolve the mysteries still 
surrounding scaling behaviors in materials such as the high-$T_c$ cuprate superconductors. 

\paragraph*{Acknowledgments}

We would like to thank Kedar Damle for suggesting a definition of spinons using valence-bond strings \cite{tang11,banerjee10}. 
The research was supported by NSFC Grant No. 11175018 (WG), the Fundamental Research Funds for the Central Universities (WG),
NSF Grant No.~DMR-1410126 (AWS), and by the Simons Foundation (AWS). HS and WG gratefully acknowledge support from the Condensed Matter Theory Visitors 
Program at Boston University and AWS is grateful to the Institute of Physics, Chinese Academy of Sciences, and Beijing Normal University for their support. 
Some of the computations were carried out using Boston University's Shared Computing Cluster.

\newpage

\baselineskip16pt

\setcounter{page}{1}
\setcounter{equation}{0}
\renewcommand{\theequation}{S\arabic{equation}}

\noindent
\centerline{\Large Supplementary Material for}
\vskip4mm

\centerline{{\Large\bf Quantum Criticality with Two Length Scales}} 
\vskip4mm

\centerline{Hui Shao, Wenan Guo, and Anders W. Sandvik}
\vskip5mm

\noindent
In {\bf Sec.~\ref{sm:crossings}} we discuss the crossing-point analysis (phenomenological renormalization) underlying the finite-size scaling studies 
summarized in Fig.~\ref{fig1}. We derive the scaling properties of the crossing points as a function of the system size and use the 2D Ising model as a 
bench-mark case to demonstrate the unbiased nature of the method when all sources of statistical errors and scaling corrections are considered.  
\vskip3mm

\noindent
In {\bf Sec.~\ref{sm:domainwalls}} we present the scaling arguments underlying the analysis of the domain-wall energy $\kappa$ of the critical $J$-$Q$ 
model in Fig.~\ref{fig3}(a). We begin with the simpler case of a generic system with discrete symmetry-breaking at a critical point with a single 
divergent length scale, deriving the scaling form of $\kappa$ in the thermodynamic limit and for finite size. We then generalize to the 
case when the thickness of the domain wall diverges faster than the correlation length and discuss the different scenarios for finite-size scaling at 
criticality. We use the 2D Ising model as an example to illustrate Monte Carlo (MC) procedures we have developed for computing the free-energy differences 
needed for $\kappa$ at thermal phase transitions. We also demonstrate conventional finite-size scaling in a classical system with a dangerously irrelevant 
perturbation; the 3D six-state clock model. For the $J$-$Q$ model at $T=0$, we discuss calculations of the ground state energy with and without domain walls
and supplement the results shown in Fig.~\ref{fig3}(a) with results for a different domain-wall boundary condition, demonstrating consistent anomalous 
scaling with the same exponent ratio $\nu/\nu'$ in both cases.

\vskip3mm

\noindent
In {\bf Sec.~\ref{sm:stiffness}}, to further motivate the two-length scaling form (\ref{aqlform2}) and its unconventional limiting behavior [demonstrated 
numerically in Figs.~\ref{fig3}(b,c)], we present derivations of the quantum-critical scaling forms of the spin stiffness and magnetic susceptibility by
generalizing the approach of Fisher {\it et al.} \cite{fisher89} to the case of two divergent length scales. 

\vskip3mm

\noindent
In {\bf Sec.~\ref{sm:temp}} we re-analyze previously published \cite{sandvik11} results for the temperature dependence of the spin correlation 
length and the magnetic susceptibility of the critical $J$-$Q$ model. We demonstrate that their $T>0$ scaling anomalies can also be explained by 
exponents modified by the exponent ratio $\nu/\nu'$.

\vskip3mm

\noindent
In {\bf Sec.~\ref{sm:qmc}} we provide some more details of the $T>0$ and $T=0$ (ground-state projector) QMC methods used in the 
studies of the $J$-$Q$ model.

\section{Crossing-point analysis}
\label{sm:crossings}

The crossing-point analysis employed in Fig.~\ref{fig1} is an extension of Fisher's ``phenomenological renormalization''. We follow essentially 
the formalism developed and tested with numerically exact transfer-matrix results for the Ising model in Ref.~\cite{fsref}, but apply it to QMC 
data. In Sec.~\ref{sm:crossings_1} we discuss formalities and derivations of the exponents governing the drifts of crossing points in the standard case, when 
there is a single divergent length scale. In Sec.~\ref{sm:crossings_2} we discuss why the single-length scaling form (\ref{aqlform1}) can still be used to 
analyze crossing points and extract the exponent $\nu$ controlling the shorter length scale, even in the case when the criticality is described by the 
two-length ansatz (\ref{aqlform2}) with the anomalous limit controlled by the longer length scale. In Sec.~\ref{sm:crossings_3} we discuss several 
practical issues and potential error sources (statistical as well as systematical) that should be properly taken into account when analyzing crossing points. 
We illustrate the procedure with data for the 2D Ising model, demonstrating the unbiased nature of the approach by reproducing the exactly known 
critical temperature and critical exponents to within small statistical errors.

\subsection{Scaling corrections and crossing points}
\label{sm:crossings_1}

Consider first the standard case of a single divergent length scale (correlation length) $\xi \propto |\delta|^{-\nu}$ as a function of the distance 
$\delta=g-g_c$ to a critical point (a classical transition driven by thermal fluctuations at $T>0$ or a quantum phase transition at $T=0$). For
some other singular quantity $A$ with the behavior $A \propto |\delta|^\kappa$ in the thermodynamic limit (valid for $g < g_c$, $g>g_c$, or both,
depending on the quantity) the finite size scaling is governed by the form
\begin{equation}
A(\delta,L) = L^{-\kappa/\nu}f(\delta L^{1/\nu}, \lambda_1 L^{-\omega_1}, \lambda_2 L^{-\omega_2}, \cdots),
\label{fullaform}
\end{equation}
where $0 < \omega_i < \omega_{i+1}$ and the variables $\lambda_i$ are irrelevant fields which in principle can be tuned by introducing some other 
interactions in the Hamiltonian \cite{omeganote}. Keeping only the most important irrelevant field, using the notation $\omega \equiv \omega_1$ for convenience, and 
suppressing the dependence on the unknown value of  $\lambda_1$, we have Eq.~(\ref{aqlform1}) in the main text. The scaling function 
is non-singular and we can Taylor expand it in the neighborhood of the critical point;
\begin{equation}
A(\delta,L) = L^{-\kappa/\nu}(a_0+a_1\delta L^{1/\nu} + b_1L^{-\omega} + \ldots).
\label{ataylor}
\end{equation}
For two system sizes $L_1=L$ and $L_2=rL$ ($r>1$), the two curves $A(\delta,L_1)$ and $A(\delta,L_2)$ take the same value (cross each other) at the point
\begin{equation}
\delta^* = \frac{a_0}{a_1}\frac{1-r^{-\kappa/\nu}}{r^{(1-\kappa)/\nu}-1}L^{-1/\nu}
+\frac{b_1}{a_1}\frac{1-r^{-\kappa/\nu-\omega}}{r^{(1-\kappa)/\nu}-1}L^{-1/\nu-\omega}.
\label{deltastar}
\end{equation}
Thus, in general the finite-size value $g^*(L)$ of the critical point defined using such curve-crossing points shifts with the system
size as $g^*(L) -g_c \equiv \delta^* \propto L^{-1/\nu}$. However, if the quantity $A$ is asymptotically size-independent at the critical point, $\kappa=0$, the
first term in Eq.~(\ref{deltastar}) vanishes and the shift is faster;
\begin{equation}
g^*(L)-g_c  \propto L^{-(1/\nu+\omega)},
\label{extraform1}
\end{equation}
where the constant of proportionality depends on the chosen aspect ratio $r$ and the generally unknown coefficients of the Taylor expansion (\ref{ataylor}).
The value of the quantity $A$ at the crossing point is obtained by inserting $\delta^*$ into Eq.~(\ref{ataylor}), which for both the general
case $\kappa\not=0$ and the special case $\kappa=0$ can be written as
\begin{equation}
A^*(L)=A(\delta^*,L) = L^{-\kappa/\nu}(a + b L^{-\omega} + \ldots),
\label{extraform2}
\end{equation}
with some constants $a$ and $b$.
Thus, in principle a crossing point analysis can be used to obtain the leading critical exponents $\kappa$ and $\nu$ as well as the subleading
exponent $\omega$. However, it should be noted that the higher-order terms in Eq.~(\ref{ataylor}) can play a significant role for system sizes
attainable in practice, and often $1/\nu + \omega$ and $\omega$ extracted from fitting to power laws according to Eqs.~(\ref{extraform1}) and
(\ref{extraform2}) should be considered only as ``effective'' exponents which change with the range of system sizes considered (with the
correct exponents obtained only for very large system sizes where the subleading corrections become negligible). To extract the critical point, 
a dimensionless quantity ($\kappa=0$) should be chosen as the convergence then is the the most rapid, given by Eq.~(\ref{extraform1}). The value 
of the critical point $g_c$ obtained from fitting to this functional form is normally not very sensitive to the imperfection of the power-law 
correction with the effective value of the exponent, as long as the fit is statistically sound.

There are many other ways of analyzing crossing points. For instance, the exponent $\nu$ can be obtained more directly than the
difficult extraction based on the correction terms in the shift analysis above. Consider a dimensionless quantity $Q$, such as the Binder ratio 
(or the corresponding cumulant). We then have, including also some terms of higher order in Eq.~(\ref{ataylor}),
\begin{equation}
Q(\delta,L)=a_0+a_1\delta L^{1/\nu} + a_2\delta^2 L^{2/\nu} + b_1L^{-\omega}  + c_{1}\delta L^{1/\nu-\omega} +  \ldots ,
\end{equation}
and from the derivative $s(\delta)$ with respect to $\delta$ or $g=g_c+\delta$ we have
\begin{equation}
s(\delta)= \frac{d Q(\delta,L)}{d\delta} = \frac{d Q(g,L)}{d g} = 
a_1L^{1/\nu}+c_{1}L^{1/\nu-\omega} + a_2\delta L^{2/\nu} + \ldots .
\label{sderiv1}
\end{equation}
We will now assume that $s(\delta)$ is positive in the region of interest, and if not we redefine it with a minus sign.
At $\delta=0$ we then have
\begin{equation}
\ln[s(0)] = c + \frac{1}{\nu} \ln(L) + d L^{-\omega} + \ldots,
\end{equation}
with some constants $c$ and $d$. Thus, for large $L$, $\ln(s)$ at the critical point depends linearly on $\ln(L)$ and the slope is 
the exponent $1/\nu$. A drawback of this method for extracting $\nu$ is that the critical point has to be determined first, and a careful
analysis should also take into account the uncertainties in the estimated value of $g_c$. 

To circumvent the requirement of having to determine $g_c$ first, we observe that, instead of evaluating the derivative (\ref{sderiv1}) exactly at the 
critical point, we can use the crossing point of the quantity $Q$ for two system sizes $(L_1,L_2)=(L,rL)$ [or, as in Ref.~\cite{fsref}, one
can use $L_2=L_1+\Delta L$ with a constant $\Delta L$, which only modifies some unimportant prefactors of the results derived below]. 
Inserting the crossing value (\ref{extraform1}) of $\delta$ into (\ref{sderiv1}) we obtain
\begin{eqnarray}
s(\delta^*,L_n)&=&a_1L_n^{1/\nu}+c_{1}L_n^{1/\nu-\omega} + a_2dL_n^{1/\nu-\omega} + \ldots \nonumber \\
&=&a_1L_n^{1/\nu}(1 + \tilde b_1L_n^{-\omega} + \ldots ),~~~~~n=1,2.
\end{eqnarray}
Having access to two different slopes at the crossing point, we can take the difference of the logarithms of these and obtain
\begin{equation}
\ln[s(\delta^*,rL)]-\ln[s(\delta^*,L)] = \frac{1}{\nu}\ln(r)+e L^{-\omega} + \ldots,
\end{equation}
with some constant $e$. We can therefore define an exponent estimate $\nu^*(L)$ corresponding to the crossing point,
\begin{equation}
\frac{1}{\nu^*(L)} = \frac{1}{\ln(r)}\ln{\left (\frac{s(\delta^*,rL)}{s(\delta^*,L)}\right )},
\label{nustar2}
\end{equation}
and this estimate approaches the correct exponent at the rate $L^{-\omega}$ for large $L$;
\begin{equation}
\frac{1}{\nu^*(L)} = \frac{1}{\nu} + g L^{-\omega} + \ldots,
\label{extraform3}
\end{equation}
with some constant $g$ and various higher-order terms again left out. 

With all the crossing-point quantities discussed above, the infinite-size values $g_c$, $Q_c$, and $1/\nu$  can be obtained by fitting data for several 
system-size pairs $(L,rL)$, using Eqs.~(\ref{extraform1}), (\ref{extraform2}), and (\ref{extraform3}). One can either use the leading form as written with 
only the asymptotically dominant correction $L^{-(1/\nu+\omega)}$ (in the case of $g_c$) or $L^{-\omega}$ (for $Q_c$ and $1/\nu$) if the system sizes are large enough for
the higher-order terms to be safely neglected, or one can include higher-order terms explicitly and fit to a larger range of system sizes. The former 
method has the advantage of the optimum fit being easier to find, while fits with multiple power laws are some times challenging or affected 
by large fluctuations in the parameters unless the statistical errors are very small.

\subsection{The case of two length scales}
\label{sm:crossings_2}

We now turn to systems with two divergent lengths, where the critical scaling is governed by Eq.~(\ref{aqlform2}). When the thermodynamic limit
corresponds to the scaling function $f(x,y)$ being a power of the first argument $x=\delta L^{1/\nu}$ for large $x$ and $y$, the effect of the 
second argument $y=\delta L^{1/\nu'}$ is the same as in the standard case of a dangerously irrelevant field scaling as $L^{-\omega'}$. 
The crossing-point analysis then remains 
the same as in the previous section. In the anomalous case, which we have termed the {\it super dangerous} perturbation, the second scaling 
argument (the longer length scale) generically controls the $L \to \infty$ behavior and demands the modified powers of $L$ in front of the
scaling function. This case requires some additional 
discussion.

In general, the scaling in this case is much more complex. In the main paper we have discussed how the correct thermodynamic limit is obtained when the 
scaling function is controlled by $y=\delta L^{1/\nu'}$. This limit corresponds directly to the intuitive physical picture of the shorter length $\xi$ 
saturating at $L^{\nu/\nu'}$ when the longer length $\xi'$ reaches $L$, and, therefore, $\xi$ should not be replaced by $L$ at criticality but instead 
by $L^{\nu/\nu'}$. This change imposes an anomalous power law at criticality for any observable which can be written as some nonzero power of the correlation 
length close to the critical point. It should be noted that, there are special non-generic observables, such as the Binder ratio, which by construction  
neither have any $L$-dependent prefactors of the finite-size scaling function $f$ nor any dependence on $\xi$ in the thermodynamic limit (e.g., the Binder ratio 
takes constant values in the phases and a different value at the critical point). In such non-generic cases there are also no modified power laws, since 
there are no powers to be modified by the ratio $\nu/\nu'$ in the first place. All other generic observables are expected to develop anomalous power laws.

We next note that, in the above large $L$ limit of $f(x,y)$, both the arguments $x=\delta L^{1/\nu}$ and $y=\delta L^{1/\nu'}$ become large. When we 
are interested in crossing points close to $\delta=0$, we are far from this limit, however. We can anticipate crossing points as in the single-length 
case when the first argument $x$ is of order one (i.e., $\delta$ is of order $L^{-1/\nu}$), 
whence the second argument is very small, $y \approx L^{1/\nu'-1/\nu} \ll 1$. There is no {\it a priori}
reason to expect that this limit is controlled by $y$. The most natural assumption, which can be tested, is that $y$ is irrelevant in this regime. 
Then we are back at a situation where the standard crossing-point analysis can be performed and the exponent delivered by such an analysis should 
generically be $\nu$, not $\nu'$. An exception is an observable which is manifestly dependent only on the longer length scale, in which case the
shorter length scale will play the role of an irrelevant correction. The simplest quantity of this kind is a length scale which is proportional to
the longer length $\xi'$ itself. In the main text we have analyzed the size $\Lambda$ of the spinon bound state and found its crossing points
to be controlled by an exponent exponent $\nu'$ which is indeed significantly larger than $\nu$, and also $\Lambda \sim L$ holds in the neighborhood
of the critical point, as expected from the scaling function controlled by $y$ when $\Lambda \sim \xi'$ in the thermodynamic limit.

We now have concluded that the limits of $f(x,y)$ when $y=\delta L^{1/\nu'} \to \infty$ and $y \to 0$ are controlled by different exponents in the generic case; 
by both $\nu$ and $\nu'$ in the former case and only by $\nu$ in the latter case. This implies an interesting cross-over behavior between these limits. In principle, 
such a cross-over can be tested explicitly by numerical data, by graphing results for a wide range of system sizes and couplings (in the case of the $J$-$Q$ 
model, that should be done inside the VBS phase) against both $\delta L^{1/\nu}$ and $\delta L^{1/\nu'}$. One should observe data collapse onto common scaling 
functions in both cases, but only in the relevant regimes controlled by the different scaling arguments; small $\delta L^{1/\nu}$ or large $\delta L^{1/\nu'}$. It would 
clearly be desirable to carry out such an analysis for the $J$-$Q$ model, 
which we have not yet done due to the large computational resources required to do this properly for sufficiently 
large system sizes. We anticipate the analysis of the cross-over to be complicated also by the small exponent $\omega$ of the leading scaling
corrections, as demonstrated in Fig.~\ref{fig1} in the main paper.

Even if no tests of the cross-overs are available currently, the two limits $y\to 0$ and $y \to \infty$ have already been confirmed in
this work; the former by the scaling of the Binder cumulant with the exponent $\nu$ (the shorter length scale) and the latter more indirectly by the 
presence of anomalous powers of $L$. An anomalous exponent which is very well converged as a function of the system size and completely inconsistent 
with any other previous scenario (neither large scaling corrections nor a first-order transition) is best provided by the domain-wall energy $\kappa$,
which is analyzed in Fig.~\ref{fig3}(a) of the main paper and also further below in Sec.~\ref{sm:domainwall_jq}.

\subsection{Tests on the 2D Ising model}
\label{sm:crossings_3}

In order to demonstrate the reliability of the method of obtaining the critical point and exponents from crossing points, and to discuss practical 
issues in implementing it, we here present results based on the Binder cumulant $U$ of the standard 2D Ising model;
\begin{equation}
U=\frac{1}{2} \left ( 3 - \frac{\langle m^4\rangle}{\langle m^2\rangle^2} \right ),
\label{isingbinder}
\end{equation}
where $m$ is the magnetization
\begin{equation}
m = \frac{1}{N}\sum_{i=1}^N \sigma_i,~~~ \sigma_i \in \lbrace -1,+1\rbrace.
\end{equation}
MC simulations were carried out on lattices of size $L\times L$ with periodic boundary conditions,
using a mix of Wulff and Swendsen-Wang (SW) cluster updates, with each sweep of Wulff updates
(where on average $\approx N$ spins are flipped) followed by an SW update where the system is decomposed into clusters, each of
which is flipped with probability $1/2$. The SW clusters are also used to measure $\langle m^2\rangle$ and $\langle m^4\rangle$ with
improved estimators (after each SW update). We carried out simulations of sizes $L=6,7,\ldots,20,22,\ldots,36,40,\ldots, 64,72,
\ldots, 128$, at $20-30$ temperatures in the neighborhood of the relevant crossing points of the Binder cumulant for system-size 
pairs $(L,2L)$, i.e., using aspect ratio $r=2$ in the expressions of Sec.~\ref{sm:crossings_1}. 
Up to $5\times 10^9$ measurements were collected for the smaller sizes and $10^8$ 
for the largest sizes. 

\begin{figure}
\center{\includegraphics[width=11cm, clip]{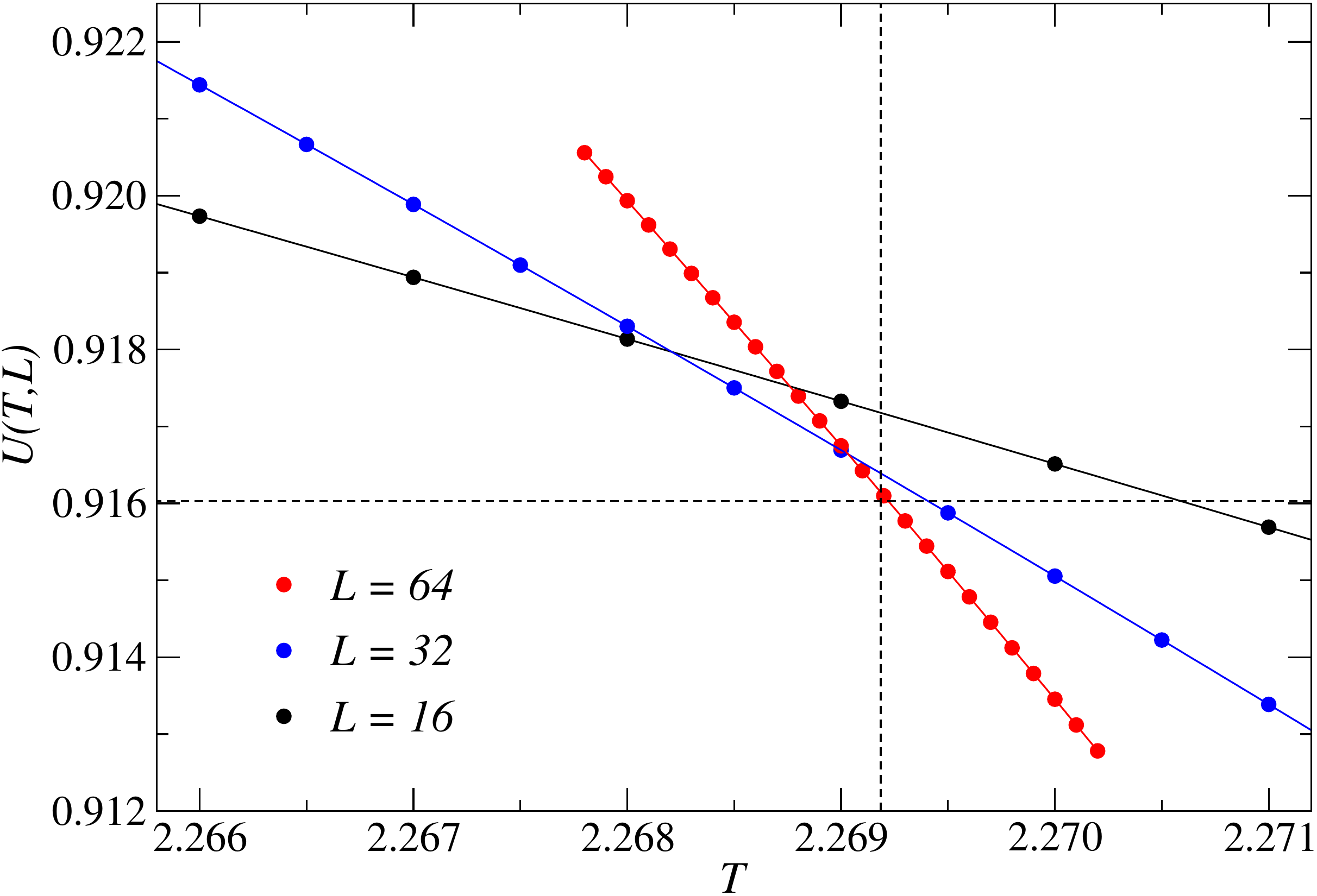}}
\vskip-0mm
\caption*{Figure S1: Binder cumulant of the 2D Ising model with $L=16,32,64$ in the neighborhood of the points at which
the curves cross each other. The vertical and horizontal dashed lines indicate the critical temperature $T_c$ and the value
of the cumulant at $T_c$, respectively. The solid curves are cubic polynomial fits to the data sets. Error bars are much smaller 
than the plot symbols.}
\vskip-1mm
\end{figure}

Figure S1 shows examples of data for three different system sizes, where cubic polynomials have been fitted to the data. The crossing
points can be extracted numerically using bisection. In order to analyze $T_c$ and $U_c$ in the thermodynamic limit, it suffices 
to consider a small number of points very close to each crossing point to be analyzed. To obtain $\nu$ from the slopes according to Eq.~(\ref{nustar2}), 
where the derivative in Eq.~(\ref{sderiv1}) is taken of the fitted polynomials, it is better to have a more extended range of points. However, 
for a very large range a high order of the polynomial has to be used in  order to obtain a good fit, and it is then better in practice to adapt 
the window size so that a relatively low order polynomial can be used. In the tests reported here, cubic polynomials were used and all fits 
were statistically sound. 

In order to compute the statistical errors (error bars) a bootstrap method can be used, i.e., by generating a large number of random samples of the 
binned MC data. Each bootstrap sample is computed using $B(L,T)$ randomly chosen bins for each system size and temperature, where $B(L,T)$ is also the total 
number of data bins available from simulations at $(L,T)$. The standard deviations of the values (the horizontal and vertical crossing points and the 
slope estimator for $1/\nu^*$) computed for these bootstrap samples correspond to the error bars, which later will be used in the fits to extrapolate 
to infinite size. In evaluating the cumulant (\ref{isingbinder}), for the full data set or a bootstrap sample, the individual expectation values 
$\langle m_i^2\rangle$ and  $\langle m_i^4\rangle$ should be computed first based on all the bins included in the sample, after which the ratio is evaluated. 
If one instead uses ratios computed for each bin separately, a statistically significant systematical error can be introduced in the ratio, due to 
nonlinear contributions to the statistical error which do not vanish as the number of bins is increased (for fixed bin size) but do decrease properly 
in the bootstrap method when the sample size is increased.

We next fit crossing points for a series of system pairs to the expected forms, Eqs.~(\ref{extraform1}), (\ref{extraform2}) with $\kappa=0$, and 
(\ref{extraform3}), and compare with exact and previous 
numerical results for the 2D Ising model. Onsager's rigorous analytical solution gives $T_c = 2\ln^{-1}(\sqrt{2}+1) \approx 2.269185314$ and $\nu=1$. 
The value of $U$ at $T_c$ is not known exactly, but Bl\"ote obtained  $U_c \approx 0.916035$ by extrapolating exact numerical finite-size
transfer-matrix data to infinite size \cite{blote93}. For the Binder cumulant the dominant subleading correction has the exponent $\omega=7/4$ 
\cite{blote93}. These results should all be obtained within statistical errors from the crossing point analysis of the MC data if sufficiently 
large systems are used and the data are analyzed using appropriate statistical methods.
For small sizes the expected higher-order corrections will cause deviations beyond the statistical errors from the 
leading-order forms, which can be detected in the goodness of the fits to the leading forms (\ref{extraform1}), (\ref{extraform2}), and (\ref{extraform3}). 
Our strategy is to remove small system sizes until a statistically sound fit is obtained for a given quantity.

The crossing points for the different size pairs $(L_i,2L_i)$, $i=1,\ldots,M$,
are not all statistically independent, because the same system size can appear in two different 
pairs. One should therefore define the goodness of the fit, $\chi^2$ per degree of freedom $N_{\rm dof}$ (the number of data points minus the number 
of parameters of the fit), with the full covariance matrix instead of just its diagonal elements (which are the conventional variances).
Using $V_i$ to denote some quantity defined based on the $(L_i,2L_i)$ crossing point (the crossing temperature $T^*$, the value of $U^*$ of $U$ 
at the crossing point, or $1/\nu^*$ obtained from the slopes evaluated using the fitted polynomial), we thus use
\begin{equation}
\chi^2 = \sum_{i=1}^M\sum_{j=1}^M (\langle V_i\rangle - V^{\rm fit}_i)[C^{-1}]^2_{ij}(\langle V_j\rangle - V^{\rm fit}_j),
\label{chi2cov}
\end{equation}
where $\langle V_i\rangle$ is either the mean value obtained from all available bins or an average obtained from the bootstrap procedure 
(the two estimates should differ only by an amount much smaller than the standard deviation based on the bootstrap analysis), $V_i^{\rm fit}$ 
is the value of the quantity evaluated using the fitted function (here a power-law correction to the infinite-size value), and $M$ is the total number 
of system-size pairs used. The covariance matrix is defined as
\begin{equation}
C_{ij} = \Bigl\langle(V_i - \langle V_i\rangle)(V_j - \langle V_j\rangle)\Bigr\rangle,~~~~~ i,j \in \{1,\ldots,M\},
\label{covarmat}
\end{equation}
where the expectation value for each pair $i,j$ for which $C_{ij} \not= 0$ is again evaluated using bootstrap sampling (as explained above for
the error bars, which correspond to the square-roots of the diagonal elements $C_{ii}$). We use of the order $100-1000$ bins and generate several 
thousand bootstrap samples to obtain accurate estimates of the covariance matrix.

To compute error bars on the extracted quantities, we repeat the fits to Eqs.~(\ref{extraform1}), (\ref{extraform2}), and (\ref{extraform3}) several 
hundred times using the bootstrap method and define the final means and statistical errors (one standard deviation) using these bootstrap samples. 
When defining $\chi^2$ as in Eq.~(\ref{chi2cov}) for data fits based on bootstrap samples, the covariance matrix (\ref{covarmat}) should be multiplied
by a factor $2$, due to the two statistically equal sources of fluctuations; the original MC fluctuations and those in the bootstrap samples.
Then, for a statistically sound fit, $\langle \chi^2\rangle/N_{\rm dof} \approx 1$ is expected for the bootstrap-averaged goodness of the fit.

To quantitatively define a criterion for an acceptable fit, we consider the standard deviation of the $\chi^2$ distribution. For $N_{\rm dof}$ degrees of 
freedom, the standard deviation of $\chi^2/N_{\rm dof}$ is $(2/N_{\rm dof})^{1/2}$. We systematically eliminate the small sizes until $\langle \chi^2\rangle/N_{\rm dof}$ 
falls roughly within two standard deviations of its expected mean;
\begin{equation}
\frac{\langle \chi^2 \rangle}{N_{\rm dof}} - 1 < 2\sqrt{\frac{2}{N_{\rm dof}}}.
\label{chi2crit}
\end{equation}
Clearly this criterion is sensitive to the quality of the data---if the elements of the covariance matrix are very small, even fits including 
only relatively large system sizes can detect the presence of higher-order corrections and not pass our test, while with noisy data also small
system sizes can be included (but the error bar on the final extrapolated value will be larger). 

If a fit satisfies the goodness-of-fit criterion (\ref{chi2crit}) it can still not be completely guaranteed that no effects of the
higher-order corrections are present in the final result, but in general one would expect any remaining systematical errors to be small
relative to the statistical error. In principle one can estimate the magnitude of the systematical error using the parameters obtained
from the fit and some knowledge or estimate of the nature of the higher-order corrections. We will not attempt to do that here, because in
general such knowledge will be very limited. To minimize possibly remaining systematical errors one can continue to exclude more system sizes
even after the soundness criterion (\ref{chi2crit}) is satisfied, at the price of increasing the statistical errors of the parameters extracted
from the fits. 

The above method implies a 'curse of good data', as less data points are actually included in the final fit when longer simulations are 
carried out for a fixed set of system sizes. However, the discarded data still contain valuable information on the convergence properties 
and can in principle be used to analyze higher-order scaling corrections (which we do not pursue here).

\begin{figure}[t]
\begin{center}
\includegraphics[width=14cm, clip]{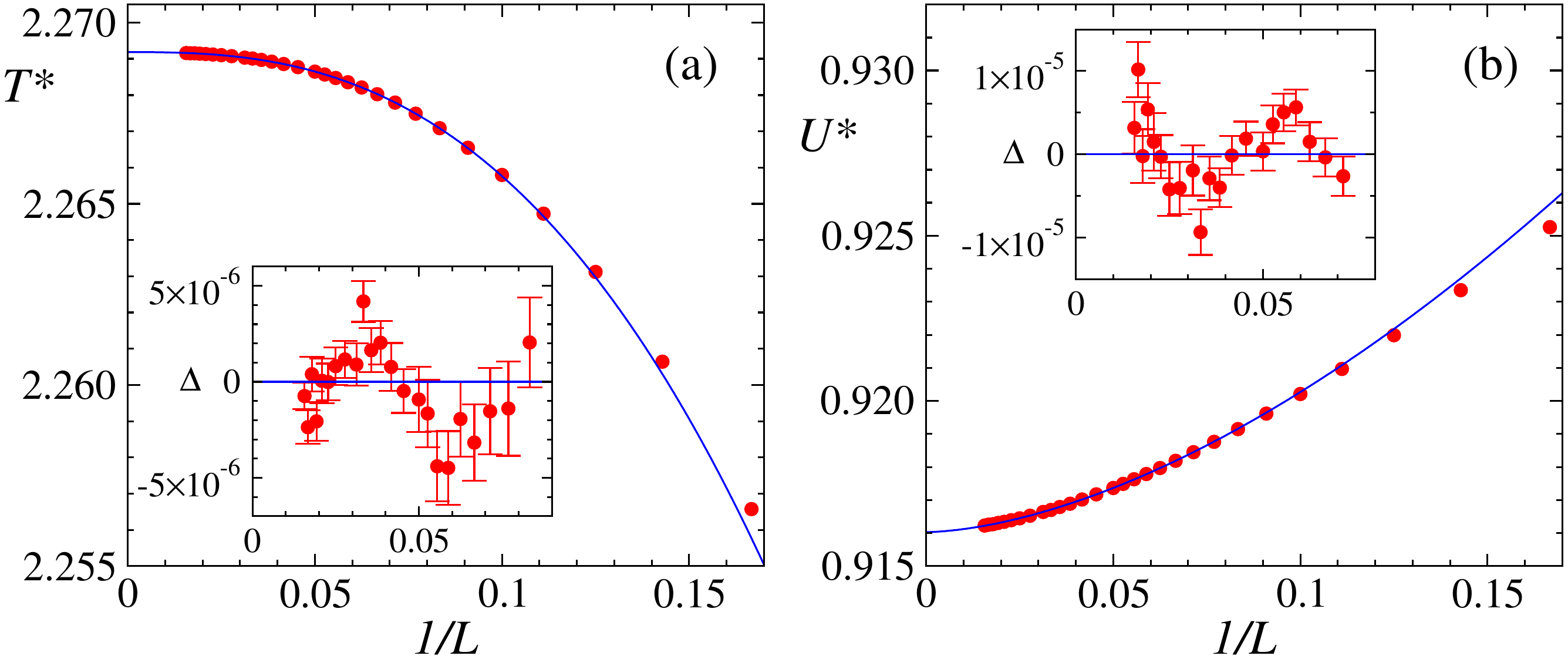}
\end{center}
\vskip-3mm
\caption*{Figure S2: Results for the 2D Ising model.
(a) Crossing temperature of the Binder cumulant for size pairs $(L,2L)$ versus $1/L$, along with a fit of the $L \ge 12$ data to the form 
(\ref{extraform1}). (b) The cumulant at the crossing points, along with a fit to the form (\ref{extraform2}) for $L \ge 14$. In both (a) and (b), 
error bars are much too small to be visible. The insets shows the difference $\Delta$ between the data and the fitted functions including the error 
bars (for only the sizes included in the fits).}
\vskip-1mm
\end{figure}

Results for the horizontal (temperature) and vertical (cumulant) crossing values of the 2D Ising model are shown in Fig.~S2. 
For the horizontal points in (a), our fits start to satisfy the criterion (\ref{chi2crit}) when including sizes $L \ge 12$  (the average goodness of 
the fit is then $\langle \chi^2\rangle/N_{\rm dof} \approx 1.6$  with $N_{\rm dof} = 20$) and we show that case in the figure. The fit gives 
$T_c = 2.2691855(5)$ and the exponent combination $1/\nu + \omega = 2.674(4)$. Thus, the critical temperature comes out correct within the remarkably small 
error bar, while $1/\nu + \omega$ is about twenty of its error bars outside the true 
(asymptotic) value $1/\nu + \omega = 2.75$. As discussed above, it is typical in finite-size scaling that corrections-to-scaling 
exponents do not come out reliably until very large systems are used, and we therefore do not consider the mismatch as a failure here, rather as a
confirmation of the known fact that the exponent should be considered as an ``effective exponent'' which slowly changes as larger
system sizes are included. 

For the crossing value of the cumulant we find a similar trend. In this case a good fit requires that
only the $L \ge 14$ points are used, giving $U_c = 0.916031(3)$ and $\omega=1.667(6)$, again with $\langle \chi^2\rangle/N_{\rm dof} \approx 1.6$ 
($N_{\rm dof} = 18$). The $U_c$ value deviates by about an error bar from Bl\"ote's result quoted above, while the correction exponent
again is relatively far (considering the size of the error bar) from its asymptotic value $\omega=1.75$. Interestingly, $1/\nu$ extracted 
as the difference of the two exponents comes out close to the correct value $1/\nu=1$, within the statistical error.

The insets of Fig.~S2 show the differences between the data points and the fitted curves. Here it can be seen that the points 
are not quite randomly distributed around $0$, as they should be if the fitted functions are of the correct form. The overall shape with noisy but
discernible minimums and maximums suggests the presence of a correction which is barely detectable for the range of system sizes at this level of statistics. 
One can then conclude that 
the deviations of $\langle \chi^2\rangle /N_{\rm dof}$ by two standard deviations from $1$ in these fits are not purely statistical fluctuations (which is not
clear from the $\langle \chi^2\rangle/N_{\rm dof}$ values alone), but due to the neglected higher-order corrections. Nevertheless, the most important extrapolated 
values $T_c$ and $U_c$ were not adversely affected statistically, thus demonstrating the ability of the effective exponent and the prefactor of 
the correction term in Eqs.~(\ref{extraform1}) and (\ref{extraform2}) to reproduce the overall trend of the data sufficiently well for extrapolating 
to infinite size. 

To illustrate the effect of excluding even more system sizes, with the minimum size $L=28$ we obtain $T_c = 2.2691831(11)$, two error 
bars away from the correct value (still a statistically acceptable match), and $U_c=0.916054(11)$, also about two error bars
from the previous (Bl\"ote's) value. From the $T_c$ fit we obtain $1/\nu +\omega = 2.70(4)$ in this case and from the $U$ fit $\omega = 1.73(5)$. 
These exponents are now correct to within statistical errors, but the error bars are about 10 times larger than before, while the error bars on $T_c$ 
and $U$ only doubled. The average value of $\langle \chi^2\rangle /N_{\rm dof}$ is very close to $1$ for both these fits and the deviations from the fitted
function look completely random. Upon excluding even more points, the error bars increase rapidly but the extracted parameters remain 
statistically in good agreement with their correct values.

\begin{figure}
\begin{center}
\includegraphics[width=10cm, clip]{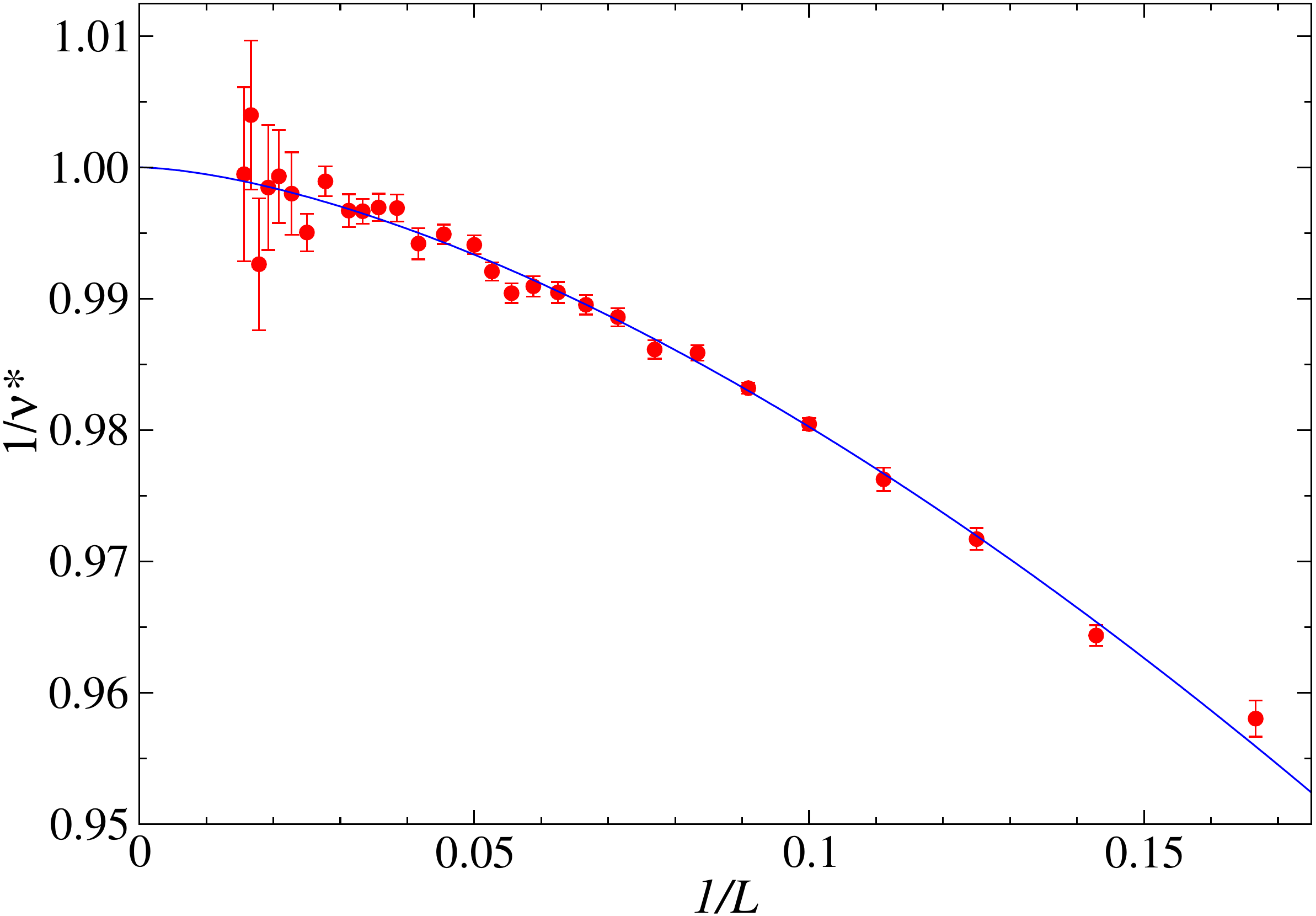}
\end{center}
\vskip-3mm
\caption*{Figure S3: Estimates of the inverse of the correlation-length exponent $\nu$ of the 2D Ising model based on the slope expression 
(\ref{nustar2}) applied to the Binder cumulant. The curve is a fit to the form (\ref{extraform1}) including all points ($L \ge 6$).}
\vskip-1mm
\end{figure}

Next, we extract the exponent $\nu$ using the log-slope formula (\ref{nustar2}). Fig.~S3 shows the results along with a fit including all the 
system sizes ($L \ge 6$). Remarkably, the fit is statistically perfect, with $\langle \chi^2\rangle/N_{\rm dof} \approx 1.0$, already at this small minimum 
size and the inverse exponent extrapolates to $1/\nu=1.0001(7)$, in excellent agreement with the exact result $1$. The slope data are much more noisy than the 
underlying $U$ values and the error bars grow very rapidly with $L$ for the largest sizes. The fit is therefore dominated by the smaller sizes. Naturally, 
the large error bars mask the effects of higher-order corrections, as discussed above. It is nevertheless remarkable that the extracted exponent $1/\nu$ does not show any effects of the neglected corrections at all, even though, again, the leading correction exponent, which comes out to $\omega=1.57(7)$, is not very close to 
the correct value $1.75$ and its error bar is large. Again, the flexibility of the leading finite-size term allows it to mimic the effects of the correction 
terms without significant effects in the extrapolation of the fit. 

It should be noted that the 2D Ising model has logarithmic corrections in addition to the higher-order scaling corrections that we have neglected
here \cite{blote93}, which is not a generic feature of critical points (except for systems at their upper critical dimension). 
The logarithms of $L$ multiply powers of $L$ higher than those of the leading 
corrections and we therefore do not expect them to affect the procedures used above.

These results demonstrate the unbiased nature of the crossing-point analysis when it is carried out properly. We have used the same scheme
to analyze the results for the $J$-$Q$ model in Fig.~\ref{fig1} of the main text. In the left column, the behavior of $\Lambda/L$ is similar to 
that of $U$ of the Ising model in Fig.~S2, with a relatively large correction exponent $\omega$ which makes the fits and extrapolations to 
$L \to \infty$ stable and and visually convincing. In the right column, it is clear that the leading correction exponent $\omega$ for $R_1$ is small, 
$\omega < 0.5$, and that there are other significant corrections present in the top two panels. The fact that the critical point nevertheless agrees perfectly 
to within small error bars with that extracted from the spinon bound state is very reassuring. As in the Ising model, the fit to $1/\nu^*$ only requires a single 
scaling correction, though it can not be excluded that this correction is an effective one, mimicking the collective effects of several corrections with the 
same sign. In any case, the extrapolations are stable, e.g., excluding some of the small-$L$ points does not dramatically change the extrapolation, 
though of course the error bar grows.

We advocate the systematic curve-crossing method as outlined above
to determine the critical temperature (or critical coupling of a quantum phase transition) and the critical exponents, instead of often used  [also 
in DQC studies \cite{sandvik07,harada13,block13}] data-collapse techniques where many choices have to be made, concerning the range of data included, 
use of corrections, etc. Although trends when increasing the system size can also be studied with data collapse [as done in Ref.~\cite{harada13})], the 
solid grounding of the present scheme directly to the finite-size scaling form (\ref{fullaform}) makes it the preferred method.

\section{Domain-wall energy}
\label{sm:domainwalls}

\begin{figure}
\begin{center}
\includegraphics[width=10cm, clip]{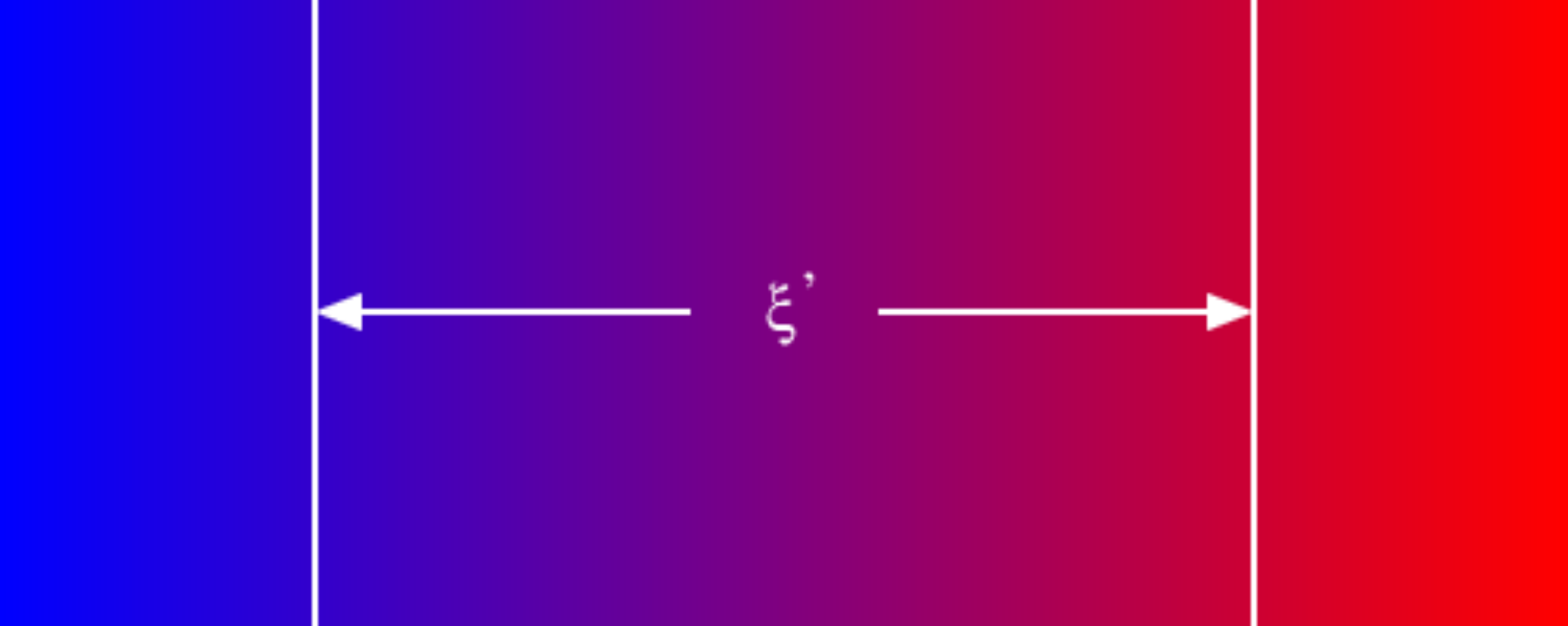}
\end{center}
\vskip-3mm
\caption*{Figure S4: A domain wall in a generic 2D system where a discrete order parameter is locked at different values (directions) to the left and right
and the twist between the two directions takes place over a region (domain-wall) of thickness $\xi'$.}
\vskip-1mm
\end{figure}

As we discussed in the main text, the fundamental longer length scale $\xi'$ in the DQC theory is the thickness of a domain wall in the VBS. In 
Fig.~S4 we illustrate a generic domain wall in a 2D system in which a discrete symmetry is broken. In the case of a broken
continuous symmetry, e.g., the magnetization vector in the XY spin model, there is no domain wall but the order parameter (its direction) gradually twists 
uniformly over the entire width $L$ of the system. This case will be discussed in Sec.~\ref{sm:stiffness} in the context of a twist of the N\'eel order 
parameter of the $J$-$Q$ model. For a discrete broken symmetry it is energetically favorable for the system to instead restrict the size of the region 
(the domain wall) over which the order parameter deviates significantly from the values imposed at the boundaries. Note, however, that the domain wall 
is not strictly fixed at some location, and, e.g., in an MC simulation the local order parameter will not detect the intrinsic width of a domain wall, 
because averaging is performed over all locations of the wall. Therefore, other means have to be employed to detect the intrinsic domain-wall thickness,
e.g., using suitably defined correlation functions.

As we showed in the main text, the length scale $\xi'$ is conveniently present in the $J$-$Q$ model in the finite-size scaling of the energy density
$\kappa$ of a VBS domain wall. Here, in Sec.~\ref{sm:domainwall_scaling} we derive the scaling form of $\kappa$, in the thermodynamic limit and for finite 
system size, using a simple Ansatz generalizing the treatment by Fisher {\it et al.} \cite{fisher89} in a different context (considered further in 
Sec.~\ref{sm:stiffness}) to the case of discrete symmetry breaking with two divergent length scales. The formalism applies both to classical and quantum 
systems. We present our MC procedures to compute $\kappa$ at classical (thermal) phase transitions, using the 2D Ising model as a concrete 
example in Sec.~\ref{sm:domainwall_ising}. We also present results for the 3D classical six-state clock model at its critical temperature
in Sec.~\ref{sm:domainwall_clock}, before describing the details of the QMC calculations of $\kappa$ for the $J$-$Q$ model at $T=0$ in 
Sec.~\ref{sm:domainwall_jq}.

\subsection{Scaling forms}
\label{sm:domainwall_scaling}

Let us first consider the case of a $d$-dimensional system with single divergent length scale $\xi \propto \delta^{-\nu}$. Following Fisher {\it et al.} 
\cite{fisher89}, we consider the singular part of the free-energy density, which we can write for a classical system at finite temperature or a quantum 
system at $T=0$ (in which case the free energy is just the ground state energy) as 
\begin{equation}
f_s(\delta,L) \propto \delta^{\nu (d+z)} Y(\xi/L) \propto \xi^{-(d+z)} Y(\xi/L),
\end{equation}
where formally the dynamic exponent $z=0$ for a classical system. Introducing a domain wall, the free-energy difference with respect to the system without domain 
wall should scale in a similar way but with a different size-dependent function \cite{fisher89};
\begin{equation}
\Delta f_s(\delta,L) \propto \xi^{-(d+z)} \tilde Y(\xi/L).
\end{equation}
This density should be understood as a quantity averaged over the inhomogeneous system (or, equivalently, in a finite system the domain wall location
is not fixed and all properties are averages over all locations of the domain wall), and the total free-energy difference is
\begin{equation}
\Delta F_s(\delta,L) \propto \xi^{-(d+z)} \tilde Y(\xi/L)L^d,
\label{dfs_a}
\end{equation}
where $L^d$ is the volume of the system.

We can also write down a different expression for the free-energy difference, by explicitly considering the cost of twisting the order
parameter. If the domain wall has width $\xi$ and the total twist of the order parameter across the wall is $\Delta\phi$, then the cost
per lattice link inside the wall is $\rho (\Delta\phi/\xi)^2$, which also defines the stiffness constant $\rho$. Outside the wall region the local
energy cost vanishes, and, since the total volume occupied by the domain wall is $\propto \xi L^{d-1}$ we have
\begin{equation}
\Delta F_s(\delta,L) \propto \rho (\Delta \phi)^2 \xi^{-1} L^{d-1}.
\label{dfs_b}
\end{equation}
Consistency in the $L$ dependence between this expression and Eq.~(\ref{dfs_a}) requires that the scaling function has the form
$\tilde Y \propto \xi/L$, and therefore
\begin{equation}
\Delta F_s(\delta,L) \propto \xi^{-(d+z-1)} L^{d-1}.
\label{dfs_c}
\end{equation}
The domain wall energy per generalized cross-section area $L^{d-1}$ of the wall (its length for $d=2$, area for $d=3$, etc.) is then
\begin{equation}
\kappa = \frac{\Delta F_s}{L^{d-1}} \propto \frac{1}{\xi^{d+z-1}},
\label{kappa_a}
\end{equation}
which no longer has any $L$ dependence and, thus, represents the behavior in the thermodynamic limit. We can also read off the 
scaling of the stiffness constant,
\begin{equation}
\rho \propto \xi^{-(d+z-2)} \propto \delta^{\nu(d+z-2)},
\end{equation}
by comparing Eqs.~(\ref{dfs_b}). and (\ref{dfs_c}).

Since we have written all expressions in terms of the correlation length, we can now switch to finite-size scaling at a critical
point by simply making the substitution $\xi \to L$. For the domain wall energy (\ref{kappa_a}) of interest here we obtain
\begin{equation}
\kappa(L) \propto L^{-(d+z-1)}.
\label{kappa_b}
\end{equation}

Now consider a system with two length scales, with a conventional correlation length $\xi \sim \delta^{-\nu}$ and a domain wall thickness 
$\xi' \sim \delta^{-\nu'}$, with $\nu' > \nu$. A simple generalization of Eq.~(\ref{dfs_a}) suggests that
\begin{equation}
\Delta F_s(\delta,L) \propto \xi^{-(d+z)} \tilde Y(\xi/L,\xi'/L)L^d.
\label{dfs_d}
\end{equation}
Note that only the shorter length scale should appear in front of the size-dependent scaling function $\tilde Y$ because the
free energy in the thermodynamic limit should only depend on the two lengths in an additive way, $f_s = a\xi^{-(d+z)} + b\xi'^{-(d+z)}$,
in order for the specific-heat exponent ($\alpha$) relation $2-\alpha = \nu(d+z)$ to hold, i.e., for hyper-scaling to apply (which we thus
assume). Since $\xi$ diverges slower than $\xi'$, $f_s$ is asymptotically dominated by the $\xi$ term, and (\ref{dfs_d}) should then describe 
the leading singular behavior.

We can also easily generalize Eq.~(\ref{dfs_b}) to a domain wall of thickness $\xi'$;
\begin{equation}
\Delta F_s(\delta,L) = \rho (\Delta\phi)^2 \xi'^{-1} L^{d-1}.
\label{dfs_e}
\end{equation}
Now consistency between Eqs.~(\ref{dfs_d}) and (\ref{dfs_e}) for both the $L$ dependence and the $\xi'$ dependence requires that
$\tilde Y \propto (L/\xi')(\xi^2/L^2)$, and we arrive at
\begin{equation}
\kappa \propto \frac{1}{\xi^{d+z-2}\xi'}
\label{kappa_c}
\end{equation}
for the scaling of $\kappa$ in the thermodynamic limit. Note the consistency of this form and the single-length form (\ref{kappa_a}) when 
$\xi' \to \xi$. In the particular case of a DQC point ($d=2$, $z=1$), Eq.~(\ref{kappa_c}) reduces to $\kappa \propto (\xi\xi')^{-1}$, which was
derived in a different way by Senthil {\it et al.}~\cite{senthil04b}.

To convert Eq.~(\ref{kappa_c}) to finite-size scaling, in the standard treatment of two length scales arising from a dangerously irrelevant 
perturbation \cite{leonard15}, the longer scale is not present in the leading finite-size scaling behavior. This can be understood physically
as follows: Upon approaching the critical point from the ordered phase, when $\xi'$ reaches $L$ we simply replace $\xi'$ by $L$. However, 
$\xi$ continues to grow and controls the scaling behavior until it reaches $L$. At the critical point also $\xi$ is replaced by $L$, and the 
critical finite-size scaling of $\kappa$ obtained from (\ref{kappa_c}) is, thus, identical to the single-length form (\ref{kappa_b}). Since
neither $\nu$ nor $\nu'$ appear here, there is no information on these exponents in the finite-size scaling of $\kappa$ in the standard
scenario. 

As we argued in the main text, there is also another possibility, namely, the growth of $\xi$ in Eq.~(\ref{kappa_c}) 
is halted when $\xi'$ reaches $L$. Then $\xi \propto L^{\nu/\nu'}$, leading to the finite-size scaling
\begin{equation}
\kappa(L) \propto L^{-1-(d+z-2)\nu/\nu'}.
\label{kappa_d}
\end{equation}
In the case of DQC, this reduces to $\kappa \propto L^{-(1+\nu/\nu')}$. It is very interesting that the ratio $\nu/\nu'$ appears here in a simple way and 
can be extracted using critical finite-size scaling. The result in Fig.~\ref{fig3}(a) leaves little doubt that $\kappa < 2$, which represents unambiguous 
evidence for anomalous scaling in the $J$-$Q$ model. Below, in Sec.~\ref{sm:domainwall_jq}, we will present details of these calculations, along with
additional results showing that $\nu/\nu' \approx 0.72$ for the $J$-$Q$ model.

\subsection{2D Ising model}
\label{sm:domainwall_ising}

\begin{figure}
\begin{center}
\includegraphics[width=5.5cm, clip]{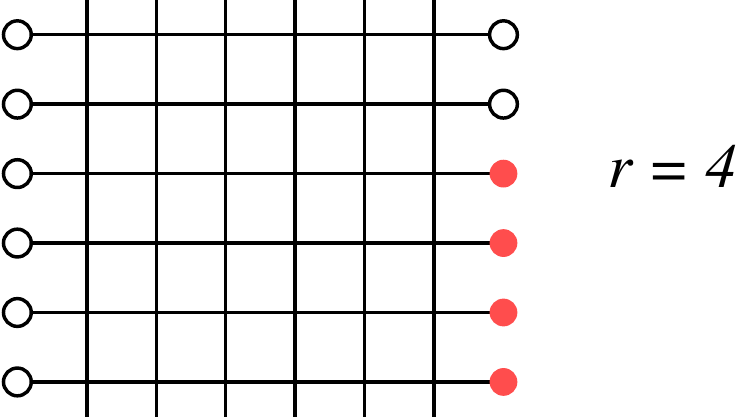}
\end{center}
\vskip-3mm
\caption*{Figure S5: Boundary conditions used to induce a domain wall in the 2D Ising ferromagnet. The black open circles and red filled circles indicate down 
and up boundary spins, respectively. The vertical location $r$ denotes the point at which the domain-wall inducing boundary is terminated. This location 
is updated, $r \to r\pm 1$, in MC updates in addition to the updates of the bulk spins. A full vertical domain wall is present when $r=L$.}
\vskip-1mm
\end{figure}

It is instructive to first test the domain-wall scaling using a simple system such as the 2D Ising model. A domain wall in the ferromagnet can be 
enforced in different ways using suitable boundary conditions. Here we use $L \times L$ systems with periodic boundaries in the $y$-direction and 
compare two different $x$ boundaries, as illustrated in Fig.~S5. The boundaries are open, with the edge columns coupled with the same 
strength $J$ as the bulk coupling to fixed spins $\sigma_i = +1$  and $\sigma_i = -1$, equivalent to boundary fields of strength $\pm J$. Here the 
domain-wall imposing column of spins to  the right extends only partially through the system, to illustrate the mechanism we use for computing 
the required free-energy difference.

It is not easy to compute the free energy in MC simulations, but it is relatively easy to compute a free-energy {\it difference}, if the two systems of 
interest, let us call them ``1'' and ``2'', can be simulated collectively as a partition function $Z_{12}=Z_1+Z_2$. If there are updates switching the 
simulation between system states 1 and 2 with detailed balance satisfied, then the free-energy difference $\Delta F_{21} = F_2-F_1 = \ln(Z_2/Z_1) = 
\ln(P_2/P_1)$, where $P_1,P_2$ are the probabilities of the simulation ``visiting'' the respective states. Such {\it multi-canonical} simulations 
\cite{marinari97} can be extended to an arbitrary number of systems $s=1,\ldots,n$, and any $\Delta F_{ij}$ can then be accessed, provided that the simulation 
can easily transition between the different states $s$.

In the studies of domain walls considered here, the different systems correspond to boundary conditions fluctuating between the normal periodic boundaries 
and the domain-wall boundaries.  To enhance the ability of the system to fluctuate between these boundary conditions of interest, the whole boundary is not 
changed at once, but in small steps where the right boundary has a change from $\sigma_i = -1$  to $\sigma_i = +1$ at some vertical location $y=r$, as
illustrated in Fig.~S5. Thus, $r=0$ corresponds to the normal periodic boundaries (no domain wall) and $r=L$ corresponds to the boundary 
enforcing a full vertical domain wall. For $0 < r < L$ the domain wall does not extend vertically through the whole system and instead has a horizontal part 
connecting to the location $y=r$ where the boundary changes. MC updates are used to move this location, $r \to r \pm 1$, using heat-bath acceptance 
probabilities. 

\begin{figure}
\begin{center}
\includegraphics[width=11cm, clip]{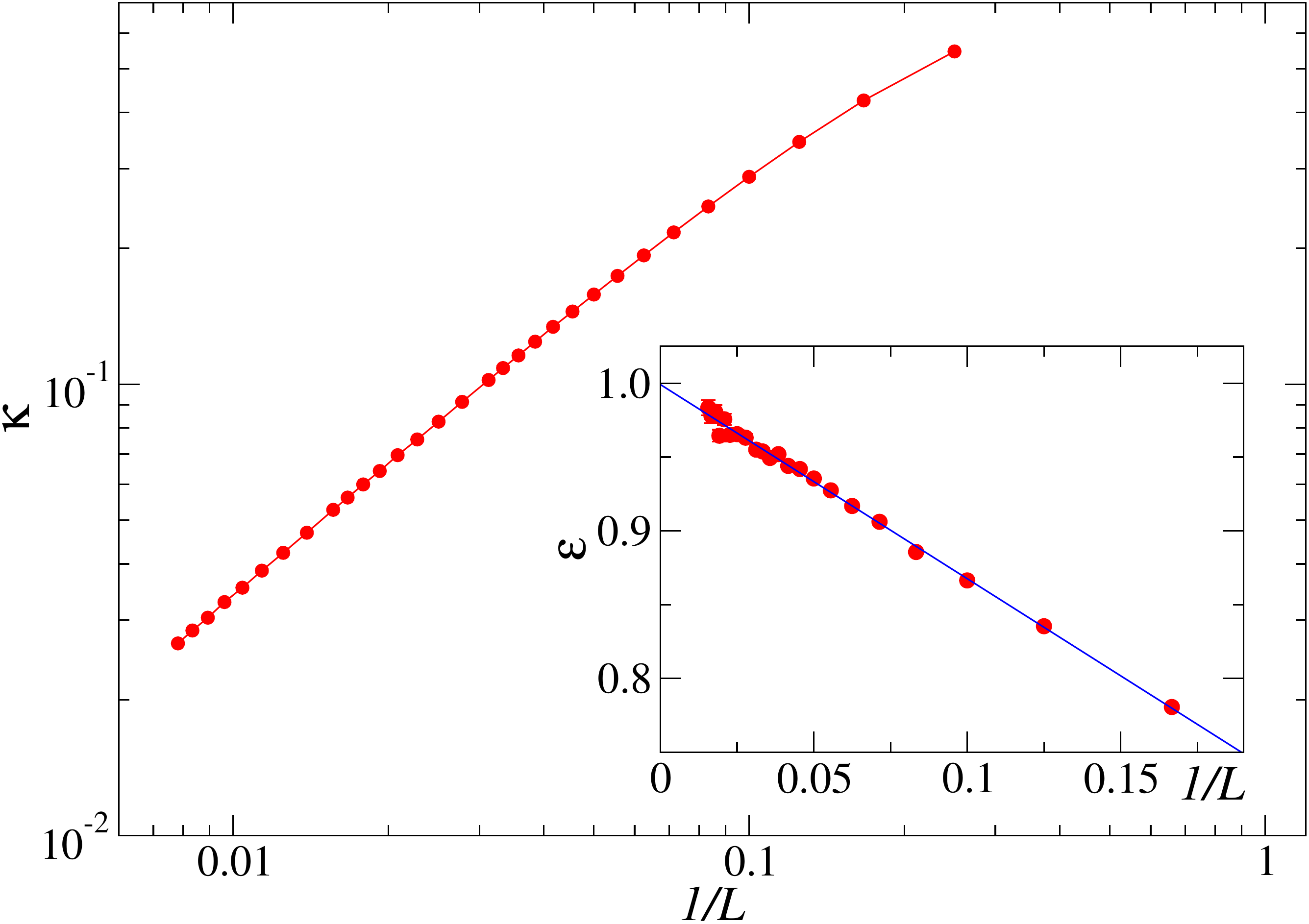}
\end{center}
\vskip-3mm
\caption*{Figure S6: Scaling of the domain-wall energy per unit length in the 2D Ising model at the critical temperature. The inset shows the running decay 
exponent obtained from data pairs $\kappa(L)$ and $\kappa(2L)$ as $\epsilon(L)=\ln[\kappa(L)/\kappa(2L)]/\ln(2)$. The results have been fitted 
to a straight line, which extrapolates to the expected value, $\epsilon \to d-1=1$, for $L \to \infty$.}
\vskip-1mm
\end{figure}

We find that the probability $P(r)$ of the boundary conditions generated is the highest, as expected, for $r=0$. There is also a local maximum
at $r=L$, and a minimum around $r=L/2$. To further increase the efficiency of the boundary moves, a weight factor $V(r)$ is multiplied with the Boltzmann 
probability for the spins and gradually adjusted until the histogram $H(r)$ of the relative number of times the boundary is at $r$ becomes almost flat. 
Then, the actual probability without the re-weighting factor is $P(r)=H(r)/V(r)$, and the free-energy difference between the systems with and without 
domain wall is (leaving out the unimportant temperature factor),
\begin{equation}
\Delta F = \ln \left (\frac{P(L)}{P(0)} \right ).
\end{equation}

MC results for $\kappa$ are shown in Fig.~S6. The inset shows the running exponent $\epsilon(L)$ extracted on the basis of size pairs $(L,2L)$
by postulating $\kappa(L)=aL^{-\epsilon(L)}$ and $\kappa(2L)=a(2L)^{-\epsilon(L)}$, whence $\epsilon(L)=\ln[\kappa(L)/\kappa(2L)]/\ln(2)$. The results are
fully compatible with $\epsilon(L) \to 1$ when $L \to \infty$, as predicted by Eq.~(\ref{kappa_b}) when $d=2,z=0$, with a correction $\propto L^{-1}$. We 
have also carried out simulations of the 3D Ising model and confirmed that $\epsilon(L) \to 2$.

\subsection{3D clock model}
\label{sm:domainwall_clock}

The existence of two length scales in the DQC theory relies heavily \cite{senthil04a,senthil04b} on an analogy with the classical 3D clock model, 
where the standard XY model is deformed by an external potential $h\cos{(q\Theta_i)}$ for all the angles $\Theta_i$. This term is known to act as a 
dangerously-irrelevant perturbation, leading to a domain-wall thickness $\xi' > \xi$. It is therefore natural to also test the scaling of the domain-wall energy 
in this case. Here we use the standard XY interaction between nearest neighbors on the 3D simple cubic lattice
\begin{equation}
H_{\rm XY} = -J \sum_{\langle ij\rangle} \cos(\Theta_i - \Theta_j),
\end{equation}
where the angles are constrained to the $q$ clock angles, $\Theta_i=n2\pi/q$, $n=0,1,\ldots,q-1$. The hard constraint is equivalent to
the limit $h/J \to \infty$ with the cosine perturbation. 

\begin{figure}
\begin{center}
\includegraphics[width=11cm, clip]{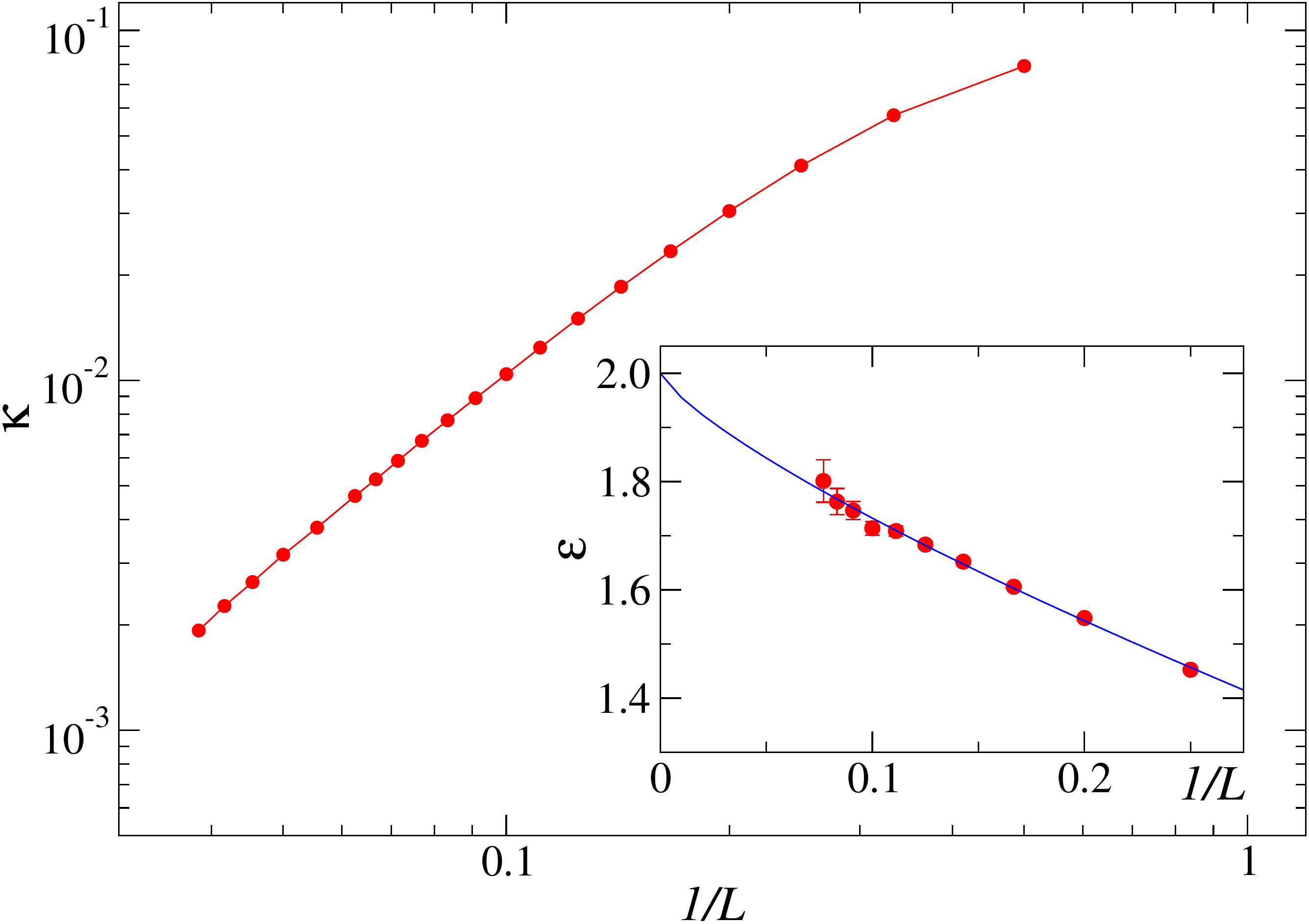}
\end{center}
\vskip-3mm
\caption*{Figure S7: Scaling of the domain-wall energy per unit area in the 3D classical $q=6$ clock model at its critical point ($T_c/J \approx 2.202$).
The inset shows the running exponent obtained from data pairs $\kappa(L)$ and $\kappa(2L)$ as $\epsilon(L)=\ln[\kappa(L)/\kappa(2L)]/\ln(2)$
and a fit to the form $\epsilon(L) = 2-aL^{-\omega}$ with $\omega \approx 0.77$.}
\vskip-2mm
\end{figure}

The exponent $\nu'$ should be independent of $h/J$ (including the fully-constrained limit considered here) but depends on $q$,
diverging as $q\to \infty$. There has been some controversy regarding methods to compute the exponent in MC simulations, as
summarized in the recent Ref.~\cite{leonard15}, but for small $q$ several calculations are nevertheless in good agreement with each other
and we can use them as reference points.

In order for the exponent ratio $\nu/\nu'$ to be significantly different from one we here use $q=6$, in which case $\nu' \approx 1.44$ and,
since the 3D XY exponent $\nu \approx 0.67$, the ratio $\nu/\nu' \approx 0.47$. Results for the domain-wall energy scaling at the
critical point are shown in Fig.~S7. The results are completely consistent with the form (\ref{kappa_b}) with $d=3,z=0$, corresponding to
the expected standard scenario where finite-size scaling is obtained from the thermodynamic-limit form by replacing both divergent length
scales by $L$. The results are completely inconsistent with the alternative scenario (\ref{kappa_d}), where the decay exponent should approach
$1+\nu/\nu' \approx 1.47$. This result reinforces the unusual form of the scaling of $\kappa$ in the $J$-$Q$ model, Fig.~\ref{fig3}(a)
of the main text, which we will discuss in more detail in the next section.

We also comment on the applicability of the generic two-length scaling form (\ref{aqlform2}) in the main paper to $\kappa$ in the clock model. Using 
the finite-size scaling we found above, we should have
\begin{equation}
\kappa(\delta,L) = L^{-2}f(\delta L^{1/\nu},\delta L^{1/\nu'}).
\end{equation}
To obtain the correct thermodynamic limit, $\kappa \to (\xi\xi')^{-1}$ when $L \to \infty$, we must have $f(x,y) \to x^\nu y^{\nu'}$,
which is also natural because, given the form in the thermodynamic limit, $f$ should be separable, $f(x,y)=f_x(x)f_y(y)$, where the 
two factors just correspond to the expected scaling forms for the length scales $\xi$ and $\xi'$ themselves. In contrast, in the $J$-$Q$ 
model we have argued for an anomalous form which corresponds to a generally non-separable scaling function with the thermodynamic limit 
controlled only by the second argument.

\subsection{J-Q model}
\label{sm:domainwall_jq}

In the $J$-$Q$ model we are interested in ground state energies of systems with and without domain walls and these can be computed in standard
QMC simulations. The multi-canonical approach employed in the previous section, developed to circumvent the difficulties of MC calculations of individual free 
energies at $T>0$, are therefore neither useful nor needed. We use the projector QMC approach  with ${\rm e}^{-\beta H}$ applied to a valence-bond trial state of 
the amplitude-product type \cite{sandvik10b,shao15}, choosing the ``projection time'' $\beta$ sufficiently large, up to $\beta=4L$, to converge the ground-state 
energy. Domain walls are introduced by boundary conditions in two different ways, schematically illustrated in Fig.~S8.

\begin{figure}[t]
\begin{center}
\includegraphics[width=10cm, clip]{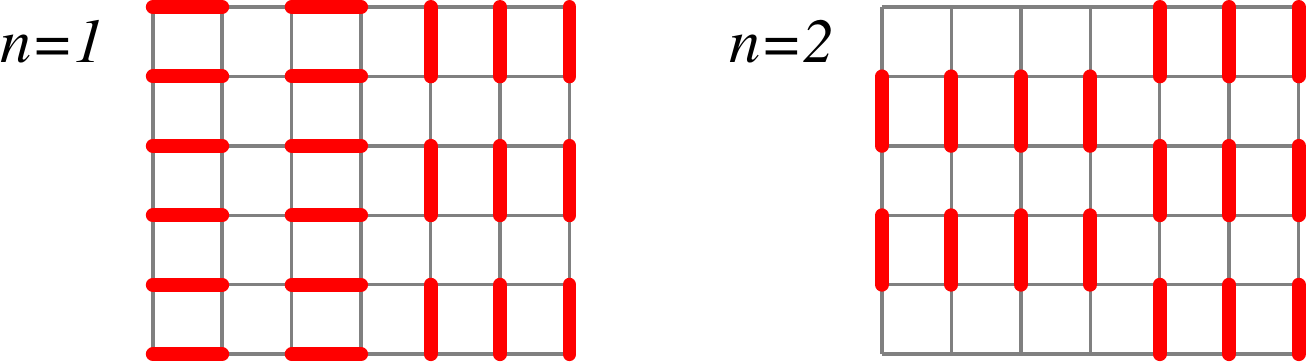}
\end{center}
\vskip-3mm
\caption*{Figure S8: Simplified pictures of VBS domain walls with total twist angle $n\pi/2$, $n=1,2$, of the order parameter between the left and right boundaries. 
In the notation introduced in the text, the boundary conditions of these two cases are denoted as $(h,v_1)$ and $(v_2,v_1)$. In QMC simulations the 
dimerization at the open $x$ boundaries is induced by weakening some of the interactions, thus explicitly breaking the symmetry between the possible 
VBS patterns. Periodic boundary conditions are employed in the $y$ direction.}
\vskip-1mm
\end{figure}

The VBS order parameter is a vector ${\bf D} = (D_x,D_y)$, where the operators corresponding to the two components can be defined as
\begin{equation}
\hat D_x = \frac{1}{N}\sum_{i=1}^N (-1)^{x_i}{\bf S}_{x_i,y_i} \cdot {\bf S}_{x_i+1,y_i},~~~
\hat D_y = \frac{1}{N}\sum_{i=1}^N (-1)^{y_i}{\bf S}_{x_i,y_i} \cdot {\bf S}_{x_i,y_i+1},
\end{equation}
where $(x_i,y_i)$ are the integer lattice coordinates of site $i$. Inside a columnar VBS phase of a large system, 
a histogram of the order parameter generated from the 
estimators of $\hat D_x$ and $\hat D_y$ in QMC simulations exhibits sharp peaks at the points $(1,0)$, $(0,1)$, $(-1,0)$, $(0,-1)$ times the magnitude 
$D$ of the order parameter. These peaks correspond to angles $m\pi/2$, $m=0,1,2,3$. As the critical point is approached, in simulations of the 
$J$-$Q$ model the histograms develop an $U(1)$ symmetry, becoming completely circular symmetric at the DQC point \cite{sandvik07,jiang08}. The length scale 
$\xi'$ controls this emergent $U(1)$ symmetry \cite{senthil04a}; upon course-graining the order parameter on length scales larger than $\xi'$ the discrete 
$Z_4$ symmetry of the VBS is apparent, while on shorter length-scales $U(1)$ symmetry develops. The thickness of a domain wall forced by suitable boundary 
conditions is controlled by this same length scale.

The four-fold symmetry of the VBS on the square lattice allows for two different types of boundary conditions, as illustrated in Fig.~S8. In the
case labeled $n=1$, the left and right sides of the lattice are forced to have VBS order with horizontal and vertical dimers, respectively, which corresponds 
to an angular difference of the order parameter $\Delta\phi=\pi/2$. In the $n=2$ graph, there is vertical dimer order at both edges, but with a relative
shift of one lattice spacing, corresponding to an angular mismatch of $\Delta\phi=\pi$. In a large system, the elementary domain wall corresponds
to $\Delta\phi=\pi/2$ and a $\pi$ wall splits into two such elementary walls. 

To compare the two cases and check for possible effects of interactions between two domain walls on the scaling of the energy, we have carried out projector 
QMC simulations with domain walls induced with total twist angles $\Delta\phi=\pi/2$ and $\pi$. Simulations without domain walls were 
carried out with similar boundary conditions, but with both the left and right walls at the same VBS angle $\phi$. The energy differences can then be computed 
without any remaining effects of edge contributions to the total energy, which for a given type of edge is the same with and without domain walls present in 
the bulk. Denoting boundary conditions enforcing horizontal dimerization at one of the edges (as in the left edge of the $n=1$ graph in Fig.~S8) by $h$ 
and vertical order with the two different phases (as shown in the $n=2$ graph) by $v_1$ and $v_2$, the systems we study with different combinations of left 
and right boundaries are $(h,h)$, $(v_1,v_1)$, $(h_,v_1)$, and $(v_2,v_1)$. The $v_1$ and $v_2$ boundaries are related by just a translation and therefore 
the edge contribution to the energy from these are the same. The domain wall contributions to the energy with the edge effects eliminated are then 
\begin{eqnarray}
\Delta E(\pi/2) & = & E(h,v_1)-[E(h,h)+E(v_1,v_1)]/2, \\
\Delta E(\pi) & = & E(v_2,v_1)-E(v_1,v_1),
\end{eqnarray}
and the corresponding size-normalized energy density is $\kappa(\Delta\phi)=\Delta E(\Delta \phi)/L$.

\begin{figure}[t]
\begin{center}
\includegraphics[width=15cm, clip]{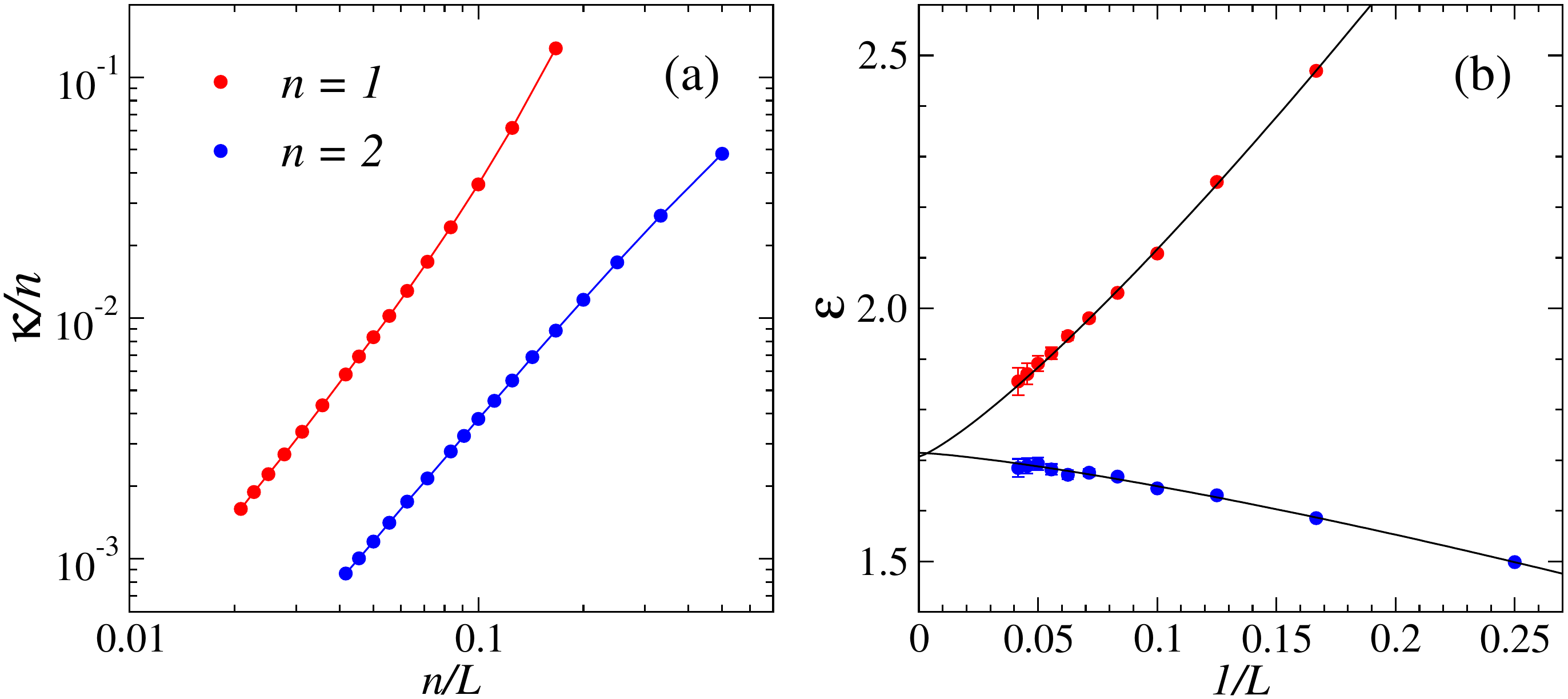}
\end{center}
\vskip-3mm
\caption*{Figure S9: (a) Domain-wall energy in the critical $J$-$Q$ model. (b) The exponent $1+\nu/\nu'$ extracted from the data in (a) as a running exponent 
(defined as in Fig.~S6) from system-size pairs ($L,2L$). Fits with power-law corrections including all data points are shown.}
\vskip-1mm
\end{figure}

QMC results for $\kappa$ computed at the estimated critical point $J/Q=0.0447$ are shown in Fig.~S9(a) [where the $\kappa=\pi$ results are
the same as those already presented in Fig.~\ref{fig3}(a)]. Here, to compare the energies on an equal footing, we divide $\kappa$ by the number $n=1,2$ 
of domain walls induced when the VBS twist angle is $\Delta\phi=n\pi/2$ and plot the results against $(L/n)^{-1}$, $L/n$ being the width over which a single 
domain wall is (on average) distributed. It is interesting, and at first sight surprising, that the $\pi/2$ domain wall is energetically much more expensive, 
since one would not expect any significant attractive interactions between the two domain walls in the $\Delta\phi=\pi$ case. We find that the lowering of 
the energy is due to enhanced fluctuations in the system with two domain walls. Recalling the emergent $U(1)$ symmetry discussed above and considering a 
$\pi/2$ domain wall between, say, boundaries at $\phi=0$ and $\phi=\pi/2$, we expect the VBS angle in the center of the system to fluctuate mainly between these 
angles.  In the case of the $\Delta\phi=\pi$ twist, there are similarly fluctuations between the angles at the edge, say $\phi=0$ and $\phi=\pi$, but here 
the system has two possible paths to go between the edges, passing either through $\phi=\pi/2$ or $\phi=-\pi/2$. Since the system is critical, there is no 
reason to expect any breaking of this symmetry.

By constructing histograms of the order parameter we have confirmed these behaviors for moderate system sizes, while for larger systems the amplitude of
the order parameter is reduced due to the critical nature of the domain walls and the histograms in both cases develop $U(1)$ symmetry. These results confirm 
that the system with $\pi$ twist is ``softer'' than that with $\Delta\phi=\pi/2$, explaining the large overall differences between the $n=1$ and $n=2$ 
results in Fig.~S9(a).

Apart from the different overall magnitudes, the power-law decay of $\kappa$ with $L$ for the largest systems is similar for $n=1$ and $2$. Fig.~S9(b) 
shows the running exponents $\epsilon(L)$ extracted from system sizes $(L,2L)$ in the same way as discussed in Sec.~\ref{sm:domainwall_ising} for the Ising 
model and Sec.~\ref{sm:domainwall_clock} for the clock model. 
The two data sets asymptotically extrapolate to the same exponent, which we have argued is $1 + \nu/\nu'$, with $\nu/\nu' = 0.715(15)$. 
The corrections are perfectly captured by a power-law term $\propto L^{-\omega}$ with the same exponent $\omega \approx 1.2 - 1.3$ for $n=1$ and $2$ but 
different signs of the  prefactor. We have also carried out calculations slightly away from the estimated critical coupling, at $J/Q=0.0450$ and $0.0445$, and 
there are no significant differences in the exponent ratio extracted at these points.

These results are key to our claims of anomalous finite-size scaling in the $J$-$Q$ model, as it is not possible to explain a non-integer decay exponent 
$\epsilon < 2$  for the domain walls within the conventional quantum-criticality scenario (as discussed above in Sec.~\ref{sm:domainwall_scaling}), 
and the results also are completely inconsistent with a first-order transition. In the latter case, VBS and N\'eel order would coexist at the transition point 
and a domain wall induced in the way explained above could possibly also be affected by coexistence inside the domain wall. However, regardless of the nature 
of the domain wall, the energy cost of the interface must scale linearly with the length of the domain wall, giving a finite $\kappa$ and a vanishing exponent 
$\epsilon(L)$  when $L \to \infty$. This seems extremely unlikely, given our data in Fig.~S9(b).

In Ref.~\cite{shao15} we employed a different approach to studying domain walls in {\it periodic} systems, by restricting the trial state used in projector 
QMC simulations in the valence-bond basis to a topological (winding number) sector corresponding to the presence of a given number of domain walls. We 
found anomalous scaling for $\kappa$, but with a somewhat larger exponent ratio $\nu/\nu'=0.80(1)$, for a different variant of the $J$-$Q$ model with products
of three singlet projectors (the $J$-$Q_3$ model) instead of the two projectors used in the model (\ref{jqham}) (the $J$-$Q_2$ model). We have also repeated 
this kind of calculation for the $J$-$Q_2$ model and again found $\nu/\nu' \approx 0.80$ for systems of small and moderate size. However, when larger
systems are considered and the statistical accuracy is sufficiently high, drifts in the exponent toward smaller values become apparent. The asymptotic behavior 
is consistent with $\nu/\nu' \approx 0.72$ obtained above with the symmetry-breaking boundaries. The previous results in Ref.~\cite{shao15} were
likely affected in the same way by remaining scaling corrections, and $\nu/\nu' \approx 0.72$ should hold universally for different variants of the
$J$-$Q$ model and for different ways of generating domain walls.

\section{Finite-size scaling of the spin stiffness and susceptibility}
\label{sm:stiffness}

In the main text we discussed the generic two-length finite-size scaling form (\ref{aqlform2}) and its different limiting behaviors compatible with 
the correct scaling of physical quantities in the thermodynamic limit. Here we discuss the behavior in the thermodynamic limit further, deriving the standard 
forms assumed in the main text for the spin stiffness $\rho_s$ and the susceptibility $\chi$ in the presence of two divergent length scales. We then argue
for the unconventional size scaling. The scaling arguments generalize similar treatments by Fisher {\it et al.}~\cite{fisher89} for a system with a single 
divergent length scale to a quantum phase transition with two divergent length scales, in a way analogous to the treatment of the domain-wall energy 
in the previous section.

The standard scenario of Fisher {\it et al.} \cite{fisher89} was formulated for interacting bosons and gives the scaling behaviors of the superfluid 
stiffness and the compressibility. The same formalism applies to a spin system as well \cite{chubukov94},  where the corresponding quantities are the 
spin stiffness $\rho_s$ and uniform magnetic susceptibility  $\chi$, which we will use in the notation here. As in Sec.~\ref{sm:domainwalls}, we again 
start from the singular part of the free-energy density,
\begin{equation}
f_s(\delta,L,\beta) \propto \delta^{\nu (d+z)} Y(\xi/L, \xi^z/\beta),
\end{equation}
where we now explicitly include the dependence on the inverse temperature $\beta$, which was assumed to be zero in the case of the quantum system 
($z>0$) in Sec.~\ref{sm:domainwalls}. In the end we will consider $\beta \to \infty$ but we will need finite $\beta$ in the derivation of the
susceptibility.

Upon imposing, by suitable boundary conditions, a total spatial phase twist $\Delta\phi$ of the continuous 
N\'eel order parameter uniformly distributed over the system, the increase in free 
energy is given by
\begin{equation}
\Delta f_s(\delta,L,\beta) = \rho_s \frac{(\Delta\phi)^2}{L^2} \propto \delta^{\nu (d+z)} \tilde Y_r(\xi/L, \xi^z/\beta).
\end{equation}
Internal consistency of this scaling form demands that $\tilde Y_r$ behaves as $(\xi/L)^2$, thus,
\begin{equation}
\rho_s \propto \xi^2 \delta^{\nu(d+z)} \propto \delta^{\nu(d+z-2)}.
\label{rhofisher}
\end{equation}
Similarly, $\chi(\Delta\phi)^2/\beta^2$ is the excess energy density needed to enforce a twist between $\tau=0$ and $\tau=\beta$ in the 
imaginary-time direction;
\begin{equation}
\Delta f_s(\delta,L,\beta) = \chi \frac{(\Delta\phi)^2}{\beta^2} \propto \delta^{\nu (d+z)} \tilde Y_\tau(\xi/L, \xi^z/\beta),
\end{equation}
where $\tilde Y_\tau$ has to behave as $(\xi^{z}/\beta)^2$. Thus, the susceptibility scales as
\begin{equation}
\chi \propto \xi^{2 z}\delta^{\nu(d+z)} \propto \delta^{\nu(d-z)}.
\label{chifisher}
\end{equation}
The finite-size scaling properties at the critical point are simply obtained from Eqs.~(\ref{rhofisher}) and (\ref{chifisher}) by replacing
$\delta^{-\nu} \propto \xi$ by the system length $L$, leading to
\begin{equation}
\rho_s \propto L^{-(d+z-2)},~~~~\chi \propto L^{-(d-z)}.
\label{finiterhochi1}
\end{equation}
In the case of $z=1$ (as in the DQC theory)  both quantities scale as $1/L$ but note that the dependence on $z$ is opposite for the two, which 
implies that the behavior seen in Fig.~\ref{fig3} in the main text can not be explained simply by $z\not = 1$.

We now generalize the above derivations to the case of two divergent length-scales, $\xi$ and $\xi'$, writing the free energy density as
\begin{equation}
f_s(\delta,L, \beta) \propto \delta^{\nu (d+z)} Y(\xi/L, \xi^z/\beta, \xi'/L, \xi'^{z}/\beta),
\end{equation}
where we have made the assumption that the same dynamic exponent governs the two time scales associated with $\xi$ and $\xi'$ (and in
principle we can generalize to two different exponents $z$ and $z'$). The excess energy due to a spatial twist is
\begin{equation}
\Delta f_s(\delta,L, \beta) = \rho_s\frac{(\Delta\phi)^2}{L^2} \propto
\delta^{\nu (d+z)} \tilde {Y}_r(\xi/L, \xi^z/\beta,\xi'/L, \xi'^{z}/\beta).
\end{equation}
Here, at first sight, there are many ways in which $\tilde {Y}_r$ can depend on its arguments in order to contain the correct $L$ dependence;
\begin{equation}
\tilde{Y}_r \propto \left (\frac{\xi}{L} \right )^{a} \left (\frac{\xi'}{L} \right )^{2-a},
\end{equation}
with arbitrary exponent $a$. However, upon approaching the critical point, when the longer length reaches $L$, we have $\xi'/L \approx 1$
and the only dependence on $L$ at that point is in the factor $(\xi/L)^a$. Thus, we can argue that $a=2$. For the thermodynamic limit we therefore
reproduce the standard results, Eq.~(\ref{rhofisher}). In a similar way we also reproduce Eq.~(\ref{chifisher}) for the susceptibility. 

For the finite-size scaling there are two physically natural options, following from two possible behaviors of the shorter length scale $\xi$ upon further 
approaching the critical point when $\xi'$ has already reached $L$: (i) $\xi$ continues to increase and eventually reaches $L$. The standard finite-size 
scaling forms (\ref{finiterhochi1}) are then again obtained by replacing $\xi \propto \delta^{-\nu}$ by $L$. (ii) The two length scales are fundamentally 
tied together, and once $\xi'$ has saturated $\xi$ is locked into its corresponding value; $\xi \propto (\xi')^{\nu/\nu'} \propto L^{\nu/\nu'}$. Making this 
replacement in the thermodynamic-limit forms (\ref{rhofisher}) and (\ref{chifisher}) leads to
\begin{equation}
\rho_s \propto L^{-(d+z-2){\nu/\nu'}},~~~~\chi \propto L^{-(d-z){\nu/\nu'}},
\label{newlscaleforms}
\end{equation}
exactly as we argued in the main text based on a direct finite-size scaling ansatz with an appropriate limit of the scaling function. As was shown 
in Fig.~\ref{fig3} in the main text, the forms (\ref{newlscaleforms}) are in excellent qualitative agreement with data for the $J$-$Q$ model, with both 
$\rho_s$ and $\chi$ decreasing slower with $L$ than in the standard forms (\ref{finiterhochi1}). Quantitative agreement is observed when using the exponent
ratio $\nu/\nu'$ extracted from the scaling of the domain-wall energy.

\section{Anomalous critical scaling at finite temperature}
\label{sm:temp}

One of the experimentally most important aspects of quantum criticality is that the quantum-critical point at $T=0$ governs the behavior 
also in a wide $T>0$ region which expands out from $(g_c,T=0)$ with increasing $T$---the so-called quantum-critical fan. In the 
standard scenario \cite{chubukov94}, the correlation length 
exactly at $g_c$ diverges when $T\to 0$ as $\xi_T \propto T^{-1/z}$ and the uniform magnetic susceptibility approaches $0$ as $\chi_T \propto T^{d/z-1}$.
These forms are not seen in simulations of the $J$-$Q$ model in the neighborhood of its critical point, however \cite{sandvik10a,sandvik11}.

Given our findings where the ratio $\nu/\nu'$ modifies the standard power laws in finite-size scaling, it is also natural to expect modifications 
of the powers of the temperature for the system in the thermodynamic limit. This expectation follows from the Euclidean path-integral mapping, where 
the inverse temperature $1/T$ of a $d$-dimensional system corresponds to the thickness in the imaginary-time dimension of the $(d+1)$-dimensional 
effective system ($L_T=c/T$, $c$ being velocity of the critical excitations). Finite-temperature scaling is therefore obtained as a generalized 
finite-size scaling in $L_T$ \cite{chubukov94}.

We here re-analyze the critical $J$-$Q$ data of Ref.~\cite{sandvik11} to test whether power laws modified by $\nu/\nu'$ can explain the observed scaling 
anomalies. The data were generated in Ref.~\cite{sandvik11} using QMC calculations on  $L\times L$ lattices with $L$ up to $512$, which allowed for
studies effectively in the thermodynamic limit down to temperatures $T/Q \approx 0.035$ ($L/L_T \gg 1$). Given that the correlation length 
diverges faster than expected and the susceptibility approaches $0$ slower than expected, in Fig.~S10 we test the forms
\begin{eqnarray}
\xi_T &\propto & T^{-1/(z\nu/\nu')}(1 + aT^{\omega_\xi}), \label{xit} \\
\chi_T &\propto & T^{(d/z-1)\nu/\nu'}(1+bT^{\omega_\chi}), \label{chit}
\end{eqnarray}
using $d=2,z=1,\nu/\nu'=0.715$ and positive correction exponents $\omega_\xi,\omega_\chi$. The correction terms reflect expected non-asymptotic contributions 
which become unimportant when $T \to 0$ but still affect the behavior for the temperatures reached in the simulations. We have multiplied $\xi_T$ by $T$ 
in Fig.~S10(a) and divided $\chi_T$ by $T$ in Fig.~S10(b), so that the results graphed versus $1/T$ should approach constants if the 
conventional forms $\xi_T \propto 1/T$ and $\chi_T \propto T$ hold. The data agree very well with the proposed anomalous forms, lending 
support to our hypothesis that finite-size anomalies carry over also to $T>0$ scaling with the same exponent ratio $\nu/\nu'$ 
modifying the power laws.

\begin{figure}[t]
\begin{center}
\includegraphics[width=14cm, clip]{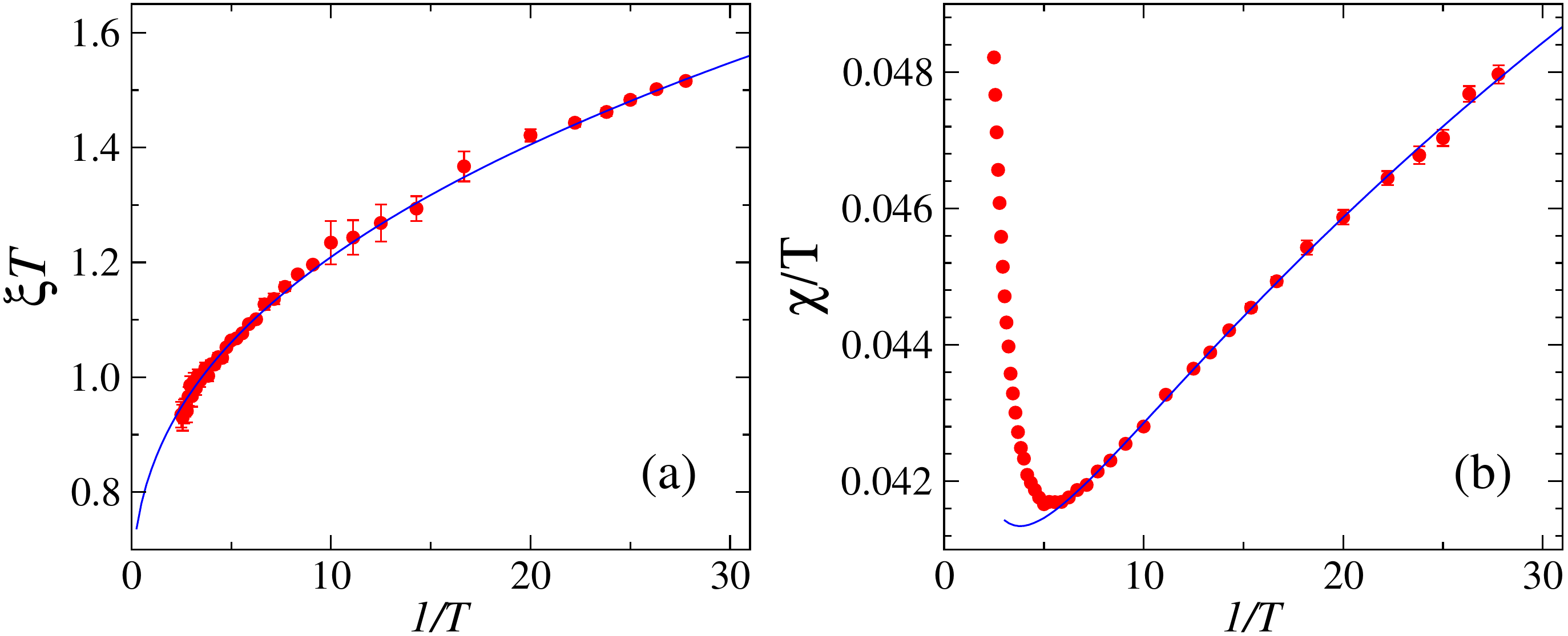}
\end{center}
\vskip-3mm
\caption*{Figure S10: Finite-temperature scaling in the critical $J$-$Q$ model based on QMC results from Ref.~\cite{sandvik11}. The
data are analyzed using the forms (\ref{xit}) and (\ref{chit}) with $d=2$, $z=1$, and the ratio $\nu/\nu'=0.715$ determined previously. 
In (a) the correlation length has been 
multiplied by $T$ and in (b) the susceptibility has been divided by $T$, so that conventional quantum-critical scaling demands 
the results to approach constants when $1/T \to \infty$. The fits shown here gave the correction exponents $\omega_\xi \approx 0.40$ 
and $\omega_{\chi} \approx 0.55$ in Eqs.~(\ref{xit}) and (\ref{chit}).}
\vskip-1mm
\end{figure}

In Ref.~\cite{sandvik11} the scaling anomaly in the correlation length was used as input in a simple picture of a deconfined gas of spinons, leading to 
quantitatively consistents relationships between numerical results for $\xi_T$, $\chi_T$, and the specific heat (the latter of which we do not analyze 
here because its anomalies are very difficult to detect). Within the spinon gas picture the susceptibility was predicted to acquire a multiplicative 
logarithmic correction. The present scenario strongly suggests a modified power law instead of a logarithm, but the consistent behaviors of the 
three quantities found in Ref.~\cite{sandvik11} still hold numerically within the temperature regime considered (since the data fits work). 

Since the spinons at the critical point
are not completely free particles in the DQC theory \cite{senthil04a,senthil04b} one cannot expect the free spinon-gas picture to remain 
strictly correct down to $T\to 0$, but it appears to apply in a window of rather low temperatures, where the log-form used for the susceptibility in 
Ref.~\cite{sandvik11} can not be distingushed from the modified power-law form proposed here. It would be interesting to carry out simulations
at still lower temperatures, to study how the logarithmic fit to $\chi_T/T$ presumably breaks down eventually and test further the anomalous 
power laws where the correction terms in Eqs.~(\ref{xit}) and (\ref{chit}) become inignificant.

\section{Quantum Monte Carlo simulations}
\label{sm:qmc}

The QMC calculations of the spin stiffness and susceptibility were carried out with the standard Stochastic Series Expansion algorithm, using the
same program as in Ref.~\cite{sandvik10a}, to which we refer for technical details and further references. For a given system size, the method produces unbiased results
only affected by well-characterized statistical errors of the MC sampling. 

Ground-state calculations in both the $S=0$ and $S=1$ sector were carried out with projector QMC simulations in the basis of valence bonds (singlet pairs) and 
unpaired spins, following Refs.~\cite{tang11,sandvik10a} and references cited there [see also Ref.~\cite{shao15}]. 
For system size $N$ and total spin $S$, there are $(N-2S)/2$ valence bonds and $2S$ unpaired spins 
with the same $z$-spin projection, i.e., the total spin-$z$ projection of the state $S^z=S$. The degrees of freedom of a bra and ket state are
importance-sampled, using the overlap of the two states as the sampling weight. This overlap is represented by a transition graph, where, in the case of the
ground state with $S=0$, the bonds form closed loops. For $S>0$ there are $2S$ ``open loops'', or strings, where for a given string the end points fall
on two unpaired spins, one in the bra and one in the ket. Such a configuration for $S=1$ is illustrated in Fig.~\ref{fig0} of the main text.

This string connecting an unpaired spin in the bra and ket states is a representation of a spinon.
The statistics of the individual strings and their cross-correlations provide information on the nature of the spinons and their collective states. 
In particular, the size of the lowest-energy $S=1$ spinon bound state in a VBS can be defined in simulations with two unpaired spins. In Ref.~\cite{tang13} 
the distance between the unpaired spins (the end points of the strings) were used for this purpose. Here we use a slightly different measure, inspired by 
the arguments of Ref.~\cite{banerjee10} for a different problem, using the entire strings in the following way \cite{damle}: Each of the two spinons, $1$, 
and $2$, in the $S=1$ state is associated with a string covering lattices sites located at $r_1(i)$, $r_2(j)$, with $i=1,\ldots, n_1$ and $j=1,\ldots ,n_2$.
We average the distance-squared $r_{ij}^2=|\vec r_1(i)-\vec r_2(j)|^2$ between two points on the two strings over all the $n_1n_2$ pairs of lattice sites
covered by the strings,
\begin{equation}
\langle r^2\rangle = \frac{1}{n_1n_2}\sum_{i=1}^{n_1}\sum_{j=1}^{n_2} r_{ij}^2,
\end{equation}
 and define the size as $\Lambda=\langle r^2\rangle^{1/2}$. We find that this definition provides a clearer signal of the spinon bound state diverging 
faster than the correlation length than definitions of $\Lambda$ based on just the unpaired spins used in Refs.~\cite{tang11,tang13}.


\begin{thebibliography}{00}
\bibitem{omeganote}
Contributions originating from the regular part of the free energy are also of the same form and in some cases, where the 
exponents of the irrelevant fields are large, they can cause the leading scaling corrections \cite{blote93}.
\end{thebibliography}
\end{document}